\documentclass[a4paper,11pt]{article}
\usepackage[affil-sl,auth-sc]{authblk}
\usepackage{amssymb,amsmath,amsbsy}
\usepackage{mathbbol}
\usepackage{bm}
\usepackage{appendix}
\usepackage{hyperref}
\usepackage[top=1.2in, bottom=1.2in,left=0.9in,includefoot]{geometry}               
\usepackage{cite}
\usepackage{float}
\usepackage{caption}
\usepackage{axodraw}
\usepackage{wasysym}
\usepackage{slashed} 
\usepackage{breqn}
\usepackage{simpler-wick}
\setkeys{breqn}{breakdepth={3}}
\usepackage[utf8]{inputenc}
\newcommand{\nc}{\newcommand} 

\newcommand{\bra}[1]{\langle #1|} \newcommand{\ket}[1]{|#1\rangle}
 \newcommand{\ie}{{\it
    i.e., }}  
 \newcommand{\bea}{\begin{eqnarray}}
\newcommand{\eea}{\end{eqnarray}} \newcommand{\nn}{\nonumber\\}
 \newcommand{\GeV}{\; \mathrm{GeV}}

\nc{\bce}{\begin{center}} \nc{\ece}{\end{center}} \nc{\be
}{\begin{equation}} \nc{\ee }{\end{equation}}
\nc{\btb}{\begin{tabular}} \nc{\etb}{\end{tabular}} \nc{\f}{\frac}
\nc{\eps}{\varepsilon} \nc{\vp}{\varphi} \def\lcal{{\cal L}}
 \nc{\tvp}{\widetilde{\varphi}} \def\wt{\widetilde}
\nc{\vpj }{\mbox{${\vp^\dag i\,\raisebox{2mm}{\boldmath
        ${}^\leftrightarrow$}\hspace{-4mm} D_\mu\,\vp}$}}
\nc{\vpjt}{\mbox{${\vp^\dag i\,\raisebox{2mm}{\boldmath
        ${}^\leftrightarrow$}\hspace{-4mm} D_\mu^{\,I}\,\vp}$}}

\usepackage[utf8]{inputenc}
\def\eq#1{eq.~(\ref{#1})}

 \def\Ref#1{ref.~\cite{#1}}
 \def\Refs#1{refs.~\cite{#1}}

\numberwithin{equation}{section}

\DeclareMathOperator{\trp}{\mathsf{T}}
\DeclareMathOperator{\cc}{\mathbb{C}}
\begin{document}

\title{\bf Feynman Rules for the Standard Model \\ Effective Field
  Theory in $R_\xi$-gauges}

\author{A.  Dedes$^{1}$\footnote{email: {\tt adedes@cc.uoi.gr}},
  ~W. Materkowska$^{2}$\footnote{email: {\tt
      Weronika.Materkowska@fuw.edu.pl}},
  ~M. Paraskevas$^{1}$\footnote{email: {\tt mparask@grads.uoi.gr}},
  ~J. Rosiek$^{2}$\footnote{email: {\tt janusz.Rosiek@fuw.edu.pl}},
  ~K. Suxho$^{1}$\footnote{email: {\tt csoutzio@cc.uoi.gr}}}
\affil{\small $^{1}$Department of Physics, University of Ioannina,
  \\ GR 45110, Ioannina, Greece}
\affil{\small $^{2}$Faculty of Physics, University of Warsaw,
  \\ Pasteura 5, 02-093 Warsaw, Poland}

\date{February 18, 2018}

\maketitle

\begin{abstract}

We assume that New Physics effects are parametrized within the
Standard Model Effective Field Theory (SMEFT) written in a complete
basis of gauge invariant operators up to dimension 6, commonly
referred to as ``Warsaw basis".
We discuss all steps necessary to obtain a consistent transition to
the spontaneously broken theory and several other important aspects,
including the BRST-invariance of the SMEFT action for linear
$R_\xi$-gauges.
The final theory is expressed in a basis characterized by SM-like
propagators for all physical and unphysical fields.
The effect of the non-renormalizable operators appears explicitly in
triple or higher multiplicity vertices.
In this \textit{mass basis} we derive the complete set of Feynman
rules, without resorting to any simplifying assumptions such as
baryon-, lepton-number or CP conservation.
As it turns out, for most SMEFT vertices the expressions are
reasonably short, with a noticeable exception of those involving 4, 5
and 6 gluons.
We have also supplemented our set of Feynman rules, given in an
appendix here, with a publicly available {\em Mathematica} code
working with the {\tt FeynRules} package and producing output which
can be integrated with other symbolic algebra or numerical codes for
automatic SMEFT amplitude calculations.
\end{abstract}

\newpage 



\tableofcontents

\newpage 

\section{Introduction and Motivation}
\label{sec:intro}

After the discovery of the Higgs boson, the picture of the Standard
Model (SM)~\cite{Weinberg:1967tq,Glashow,Salam} being a spontaneously
broken gauge theory at an Electroweak Scale (EW) with $v\sim 246 \GeV$
has been theoretically established and experimentally confirmed to a
significant accuracy. Nevertheless, new physics beyond the SM may be
hidden in the experimental errors of measurements that are becoming
increasingly accurate at the LHC. Such phenomena can be parametrized
in terms of the so-called SM Effective Field Theory
(SMEFT)~\cite{Weinberg:1980wa, Coleman:1969sm,
  Callan:1969sn}\footnote{For reviews
  see~\Refs{Manohar:1996cq,Willenbrock:2014bja,Passarino:2016pzb}.},
where, assuming $\Lambda$ to be the typical energy scale of the SM
extension, the observable effects are suppressed by powers of the
expansion parameter $v/\Lambda$.  The SM's weak response to a more
fundamental theory (effective or not) living at $\Lambda$ may be due
to the fact that such a scale is far above the EW scale \ie $\Lambda
\gg v$, or because non-renormalizable, UV-dependent couplings are,
somehow, small.

Besides the verification of the SM gauge group and content, a renewed
interest in the SMEFT arises from the fairly recent completion of all
gauge invariant, independent, (mass) dimension-6 operators, first
conducted in a study by Buchm\"uller and
Wyler~\cite{Buchmuller:1985jz} in 1985 and lately amended by the
Warsaw university group~\cite{Grzadkowski:2010es} in 2010.  We shall
refer to this set of operators as the ``Warsaw'' basis. In this basis
there are 59+1 baryon-number conserving\footnote{In counting, we
  include the lepton-number $d=5$ violating
  operator~\cite{Weinberg:1979sa} but do not count hermitian
  conjugated operators and suppress fermion flavor dependence.} and 4
baryon-number violating operators.

If physics beyond the SM lies not too far from the EW scale, so that
is invisible, but also not too close to the EW scale, so that the
effective field theory description (EFT) does not fail, then SMEFT
observables should encode possible deviations from the SM to order
$(v/\Lambda)^2$ no matter what the fundamental (UV) theory is. A
serious attempt in calculating such observables should start by first
writing down the Feynman rules for propagators and vertices for
physical fields, after spontaneous symmetry breaking (SSB) of the
effective theory, in a way that consistently renders the theory
renormalizable in the ``modern'' sense - here of absorbing infinities
into a finite number of counterterms up to order $(v/\Lambda)^2$.  One
major criterion for this to be realized is that the gauge boson
propagators vanish for momenta $p\to \infty$ as $p^{-2}$ so that the
theory satisfies usual power counting rules for renormalizability, as
in the SM for example.
In 1971, 't Hooft~\cite{tHooft:1971qjg} and B. Lee~\cite{Lee:1971kj}
showed that this can be realized in a linear gauge which a year later
extended to a larger class of renormalizable gauges by Fujikawa, Lee,
Sanda~\cite{Fujikawa:1972fe}, and Yao~\cite{Yao:1973am}.  This class
of renormalizable gauges, called $R_\xi$-gauges, can be parametrized
by one or more arbitrary constants, collectively written as $\xi$.
In addition to the smooth behavior of the propagators, $R_\xi$-gauges
allow for eliminating ``unwanted'' mixed terms between physical gauge
bosons and unphysical (Goldstone) scalar fields in spontaneously
broken gauge theories.

To the best of our knowledge, quantization of SMEFT in linear
$R_\xi$-gauges does not exist in the literature thus far.  What
complicates the picture of quantization in $R_\xi$-gauges, or as a
matter of fact in every other class of gauges, is twofold: a) field
redefinitions and reparametrizations and b) mixed field strength
operators. A careful treatment of the former to retain gauge
invariance is necessary~\cite{Passarino:2016saj} while properly
rotating away (but not completely eliminating from vertices) the
latter, results in SM-like propagators for physical and unphysical
fields.  More specifically, in this paper we consider SSB of the
``Warsaw" basis theory and present a full set of Feynman rules in
$R_\xi$-gauge in a \textit{mass basis}, with the following features:
\begin{itemize} 
\item No restriction is made for the structure of flavor violating
  terms and for CP-, lepton- or baryon-number conservation,
\item SMEFT is quantized in $R_\xi$-gauges written with four different
  arbitrary gauge parameters, $\xi_\gamma, \xi_Z, \xi_W, \xi_G$ for
  better cross checks of physical amplitudes.
\item Gauge fixing and ghost part of the Lagrangian is chosen to be
  SM-like and preserve Becchi, Rouet, Stora~\cite{Becchi:1975nq}, and
  Tyutin~\cite{Iofa:1976je} (BRST) invariance.
\item All bilinear terms in the Lagrangian have canonical form, both
  for physical and unphysical Goldstone and ghost fields; all
  propagators are diagonal and SM-like.
\item Feynman rules for interactions are expressed in terms of
  physical SM fields and canonical Goldstone and ghost fields.
\end{itemize}

We are aware that in the literature there are many calculations done
already within SMEFT, including several articles with loop
calculations usually performed in unitary or non-linear gauges, see
for example \Ref{Passarino:2016pzb} and references therein.  However,
we think that a full set of Feynman rules written (and coded in the
symbolic computer program) in the $R_\xi$-gauges, including in
addition the most general structure of the flavor violating terms, is
something that can largely simplify further such analyses.
Especially, having such collection is useful because the number of
primary vertices in SMEFT in $R_\xi$-gauges is huge: 380 without
counting the hermitian conjugates (surprisingly, for most SMEFT
vertices the Feynman rules are reasonably short, with an exception of
self-interactions of 4, 5 and 6 gluons).  An explicit diagrammatic
representation for all interaction vertices will minimize possible
mistakes that arise from missing terms or even entire diagrams in
amplitude calculations.  Furthermore, implementation of them as a
``model file'' to the {\tt FeynRules} package~\cite{Alloul:2013bka}
produces an output ready to be further used in symbolic or numeric
programs for amplitude calculations.

The procedure we followed in deriving the SMEFT Feynman rules consists
of the following steps\footnote{Steps 1 and 2 have been discussed in
  numerous earlier papers e.g., \Ref{Alonso:2013hga}, but we include
  them here for completeness and consistency.}: 
\begin{enumerate}
\item within the ``Warsaw'' basis, given for reference in
  section~\ref{sec:SM6}, we perform the SSB mechanism and further
  field and coupling rescalings with constant parameters which have no
  effect on the $S$-matrix elements (up to $\mathcal{O}(\Lambda^{-3})$
  corrections). They make all bilinear terms of gauge, Higgs and
  fermion fields canonical [section~\ref{sec:eigen}],
\item we discuss ``oblique'' corrections to the SM vertices, coming
  from the constant field and coupling redefinitions when moving from
  weak to mass basis [section~\ref{sec:SMcoup}],
\item we introduce suitable $R_\xi$-gauge fixing and ghost terms in
  the Lagrangian, in a way that renders also the ghost propagators
  diagonal. The new terms eliminate the ``unwanted" gauge-Goldstone
  mixing and establish BRST invariance.  Thus, in the mass basis of
  SMEFT all quadratic terms of physical (SM particles) and unphysical
  (Goldstone bosons and ghosts) become SM-like
  [section~\ref{sec:gaugefix}],
\item we evaluate Feynman rules for all sectors of the theory in
  $R_\xi$-gauges.  [Appendix~\ref{app:vert}].
 \end{enumerate}
Then, in section~\ref{sec:feyrul} we describe the features of the {\tt
SmeftFR} package generating automatically relevant Feynman rules in
various formats which could be used for automatised symbolic
calculations of transition amplitudes or imported to Monte Carlo
generators. We conclude in section~\ref{sec:Con}.

\section{Notation and conventions for the SMEFT Lagrangian}
\label{sec:SM6}
\begin{table}[t]
\renewcommand{\arraystretch}{1.5} \btb{|c|c|c|c|c|c|c|}
\hline \multicolumn{1}{|c|}{$ $} & \multicolumn{5}{|c|}{fermions} &
\multicolumn{1}{|c|}{scalars}\\ \hline
\renewcommand{\arraystretch}{1.} $\begin{array}{c} \\ \text{field}
  \\ \\
\end{array}$ &
\renewcommand{\arraystretch}{1.}$l^{\prime
  j}_{Lp}=\left(\begin{array}{c} \nu_{Lp}^\prime \\
e_{Lp}^\prime \end{array}\right) $ & $e^\prime_{Rp}$ &
\renewcommand{\arraystretch}{1.}$q^{\prime \alpha j}_{Lp} = \left
( \begin{array}{c} u_{Lp}^{\prime \,\alpha} \\
d_{Lp}^{\prime\, \alpha} \end{array} \right )$ &
$u^{\prime\alpha}_{Rp}$ & $d^{\prime\alpha}_{Rp}$ &
\renewcommand{\arraystretch}{1.}  $\vp^j=\left ( \begin{array}{c}
  \vp^+ \\ \vp^0 \end{array} \right )$\\
\hline
hypercharge $Y$ & $-\f12$ & $-1$ & $\f16$ & $\f23$ & $-\f13$ &
$\f12$\\
\hline \etb
\caption{\sf The SM matter content in the gauge basis.  Isospin,
  colour and generation indices are indicated with $j=1,2$,
  $\alpha=1...3$ and $p=1...3$, respectively.  \label{tab:matter}}
\end{table}

Throughout this article we use the notation and conventions of
\Ref{Grzadkowski:2010es}.  However, in order to distinguish between
the fields and parameters of the initial, gauge basis and the final,
mass basis, we use \textit{primed} notation for fermion fields and
their Wilson coefficients in the former, reserving the
\textit{``unprimed''} symbols for the physical mass eigenstates basis,
where flavor space rotations have been performed.
In addition, and not to clutter the notation further as compared to
\Ref{Grzadkowski:2010es}, we absorb the theory cut-off scale $\Lambda$
in the definitions of Wilson coefficients, rescaling them
appropriately as $C^{(5)}_X / \Lambda \to C^{(5)}_X$, $C^{(6)}_X /
\Lambda^2 \to C^{(6)}_X$.

For completeness and reference, in Tables~\ref{tab:no4ferm} and
~\ref{tab:4ferm} we list all, gauge independent, dimension-6 operators
of the ``Warsaw'' basis derived in \Ref{Grzadkowski:2010es}. The only
dimension-5 operator, the lepton-number violating
operator~\cite{Weinberg:1979sa}, reads
\be
\label{qnunu}
Q_{\nu\nu} = \eps_{jk} \eps_{mn} \vp^j \vp^m (l^{'k}_{Lp})^T \,
\mathbb{C}\, l^{'n}_{Lr} ~\equiv~ (\tvp^\dag l'_{Lp})^T\, \mathbb{C}\,
(\tvp^\dag l'_{Lr})\;,
\ee
where $\mathbb{C}$ is the charge conjugation matrix in notation of
\Ref{Grzadkowski:2010es}. Then the full gauge invariant Lagrangian, up
to $\mathcal{O}(\Lambda^{-3})$ corrections, takes the form
\be
\lcal = \lcal_{\mathrm SM}^{(4)} + C^{\nu\nu} Q_{\nu\nu}^{(5)} +
\sum_{X} C^{X} Q_X^{(6)}+ \sum_{f} C^{'f} Q_f^{(6)} \;,
\label{Leff}
\ee
where $Q_X^{(6)}$ denotes dimension-6 operators that do not involve
fermion fields, \ie operators entitled as $X^3, \varphi^6, \varphi^4
D^2, X^2 \varphi^2$ columns of Table~\ref{tab:no4ferm}, while
$Q_f^{(6)}$ denotes operators that contain fermion fields among other
fields \ie all other operators in Tables~\ref{tab:no4ferm} and
~\ref{tab:4ferm}. The renormalizable part of the Lagrangian is (we
suppress generation indices here),
\begin{align}
\mathcal{L}_{SM}^{(4)} & = -{1\over 4} G_{\mu\nu}^A
G^{A\mu\nu}-{1\over 4} W_{\mu\nu}^I W^{I\mu\nu}-{1\over 4} B_{\mu\nu}
B^{\mu\nu} +( D_\mu \varphi)^\dag ( D^\mu \varphi) + m^2 \varphi^\dag
\varphi -{1\over 2}\lambda (\varphi^\dag \varphi)^2 \nonumber\\
& +\ i(\bar l^\prime_L \slashed{D} l^\prime_L + \bar e^\prime_R
\slashed{D} e^\prime_R + q^\prime_L \slashed{D} q^\prime_L + \bar
u^\prime_R \slashed{D} u^\prime_R +\bar d^\prime_R \slashed{D}
d^\prime_R)\nn
& -\ (\bar l^\prime_L \Gamma_e e^\prime_R \varphi + \bar q^\prime_L
\Gamma_u u^\prime_R \tvp + \bar q^\prime_L \Gamma_d d^\prime_R \varphi
+ \mathrm{H.c.})\;.  \label{eq:smeft}
\end{align}
As compared to \Ref{Grzadkowski:2010es} we slightly change the
notation for the gauge group generators while keeping all other
conventions identical. The covariant derivative then reads,
\bea
{D}_\mu = \partial_\mu + i {g'} B_\mu Y+i {g} W_\mu^I T^I + i g_s
G_\mu^A \mathcal{T}^A\;,
\eea
where the weak hypercharge $Y$ assigned to the fields is given in
Table~\ref{tab:matter}. In fundamental representation, the generators
for $SU(2)$ read $T^I=\tau^I/2$ with $\tau^I$ ($I$=1,2,3) being the
Pauli matrices and for $SU(3)$ read $\mathcal{T}^A=\lambda^A/2$ with
$\lambda^A$ ($A$=1,$\dots$,8) being the Gell-Mann matrices. 
The Hermitian derivatives appearing in $\psi^2 \varphi^2 D$ class of Table~\ref{tab:no4ferm} are defined as, 
\bea
\vpj \equiv i\mbox{${\vp^\dag \,(D_\mu -\raisebox{2mm}{\boldmath
        ${}^\leftarrow$}\hspace{-4mm} D_\mu)\, \vp}$}~,~~  \vpjt \equiv i\mbox{${\vp^\dag \,(\tau^I D_\mu -\raisebox{2mm}{\boldmath
        ${}^\leftarrow$}\hspace{-4mm} D_\mu \tau^I )\, \vp}$}\,,
\eea 
with $\mbox{${\vp^\dag \raisebox{2mm}{\boldmath
        ${}^\leftarrow$}\hspace{-4mm} D_\mu\,\vp}$}\equiv (D_\mu \vp )^\dag \vp$. 
The field strength tensors are,
\bea
G_{\mu\nu}^A&=& \partial_\mu G_\nu^A - \partial_\nu G_\mu^A-
g_sf^{ABC} G_\mu^B G_\nu^C\,,\\
W_{\mu\nu}^I&=& \partial_\mu W_\nu^I - \partial_\nu W_\mu^I
-g\epsilon^{IJK} W_\mu^J W_\nu^K\,, \\
B_{\mu\nu}&=&\partial_\mu B_\nu - \partial_\nu B_\mu \, ,
\eea
and their dual tensors read $\widetilde{X}_{\mu\nu}={1\over 2} \epsilon_{\mu\nu\rho\sigma}X^{\rho\sigma} \,(\epsilon_{0123}=+1)$, with $X\equiv B,W^I$ or $G^A$.
Finally, we consider the SMEFT accurate up to
$\mathcal{O}(\Lambda^{-3})$ corrections and therefore all relations
obtained within it are accurate up to this level of approximation. We
will implicitly make use of this property in our derivations without
making any further notice.

\begin{table}[t] 
\centering \renewcommand{\arraystretch}{1.5} \btb{||c|c||c|c||c|c||}
\hline \hline \multicolumn{2}{||c||}{$X^3$} &
\multicolumn{2}{|c||}{$\vp^6$~ and~ $\vp^4 D^2$} &
\multicolumn{2}{|c||}{$\psi^2\vp^3$}\\ \hline
$Q_G$ & $f^{ABC} G_\mu^{A\nu} G_\nu^{B\rho} G_\rho^{C\mu} $ & $Q_\vp$
& $(\vp^\dag\vp)^3$ & $Q_{e\vp}$ & $(\vp^\dag \vp)(\bar l'_p e'_r
\vp)$\\
$Q_{\wt G}$ & $f^{ABC} \wt G_\mu^{A\nu} G_\nu^{B\rho} G_\rho^{C\mu} $
& $Q_{\vp\Box}$ & $(\vp^\dag \vp)\raisebox{-.5mm}{$\Box$}(\vp^\dag
\vp)$ & $Q_{u\vp}$ & $(\vp^\dag \vp)(\bar q'_p u'_r \tvp)$\\
$Q_W$ & $\eps^{IJK} W_\mu^{I\nu} W_\nu^{J\rho} W_\rho^{K\mu}$ &
$Q_{\vp D}$ & $\left(\vp^\dag D^\mu\vp\right)^* \left(\vp^\dag
D_\mu\vp\right)$ & $Q_{d\vp}$ & $(\vp^\dag \vp)(\bar q'_p d'_r \vp)$\\
$Q_{\wt W}$ & $\eps^{IJK} \wt W_\mu^{I\nu} W_\nu^{J\rho}
W_\rho^{K\mu}$ &&&&\\
\hline \hline \multicolumn{2}{||c||}{$X^2\vp^2$} &
\multicolumn{2}{|c||}{$\psi^2 X\vp$} &
\multicolumn{2}{|c||}{$\psi^2\vp^2 D$}\\ \hline $Q_{\vp G}$ &
$\vp^\dag \vp\, G^A_{\mu\nu} G^{A\mu\nu}$ & $Q_{eW}$ & $(\bar l'_p
\sigma^{\mu\nu} e'_r) \tau^I \vp W_{\mu\nu}^I$ & $Q_{\vp l}^{(1)}$ &
$(\vpj)(\bar l'_p \gamma^\mu l'_r)$\\
$Q_{\vp\wt G}$ & $\vp^\dag \vp\, \wt G^A_{\mu\nu} G^{A\mu\nu}$ &
$Q_{eB}$ & $(\bar l'_p \sigma^{\mu\nu} e'_r) \vp B_{\mu\nu}$ & $Q_{\vp
  l}^{(3)}$ & $(\vpjt)(\bar l'_p \tau^I \gamma^\mu l'_r)$\\
$Q_{\vp W}$ & $\vp^\dag \vp\, W^I_{\mu\nu} W^{I\mu\nu}$ & $Q_{uG}$ &
$(\bar q^\prime_p \sigma^{\mu\nu} \mathcal{T}^A u^\prime_r) \tvp\,
G_{\mu\nu}^A$ & $Q_{\vp e}$ & $(\vpj)(\bar e'_p \gamma^\mu e'_r)$\\
$Q_{\vp\wt W}$ & $\vp^\dag \vp\, \wt W^I_{\mu\nu} W^{I\mu\nu}$ &
$Q_{uW}$ & $(\bar q'_p \sigma^{\mu\nu} u'_r) \tau^I \tvp\,
W_{\mu\nu}^I$ & $Q_{\vp q}^{(1)}$ & $(\vpj)(\bar q'_p \gamma^\mu
q'_r)$\\
$Q_{\vp B}$ & $ \vp^\dag \vp\, B_{\mu\nu} B^{\mu\nu}$ & $Q_{uB}$ &
$(\bar q'_p \sigma^{\mu\nu} u'_r) \tvp\, B_{\mu\nu}$& $Q_{\vp
  q}^{(3)}$ & $(\vpjt)(\bar q'_p \tau^I \gamma^\mu q'_r)$\\
$Q_{\vp\wt B}$ & $\vp^\dag \vp\, \wt B_{\mu\nu} B^{\mu\nu}$ & $Q_{dG}$
& $(\bar q'_p \sigma^{\mu\nu} \mathcal{T}^A d'_r) \vp\, G_{\mu\nu}^A$
& $Q_{\vp u}$ & $(\vpj)(\bar u'_p \gamma^\mu u'_r)$\\
$Q_{\vp WB}$ & $ \vp^\dag \tau^I \vp\, W^I_{\mu\nu} B^{\mu\nu}$ &
$Q_{dW}$ & $(\bar q'_p \sigma^{\mu\nu} d'_r) \tau^I \vp\,
W_{\mu\nu}^I$ & $Q_{\vp d}$ & $(\vpj)(\bar d'_p \gamma^\mu d'_r)$\\
$Q_{\vp\wt WB}$ & $\vp^\dag \tau^I \vp\, \wt W^I_{\mu\nu} B^{\mu\nu}$
& $Q_{dB}$ & $(\bar q'_p \sigma^{\mu\nu} d'_r) \vp\, B_{\mu\nu}$ &
$Q_{\vp u d}$ & $i(\tvp^\dag D_\mu \vp)(\bar u'_p \gamma^\mu
d'_r)$\\ \hline \hline \etb
\caption{\sf Dimension-6 operators other than the four-fermion ones
  (from \Ref{Grzadkowski:2010es}). For brevity we suppress fermion
  chiral indices $L,R$. \label{tab:no4ferm}}
\end{table}

\begin{table}[t]
\centering \renewcommand{\arraystretch}{1.5}
\begin{tabular}{||c|c||c|c||c|c||}
\hline\hline \multicolumn{2}{||c||}{$(\bar LL)(\bar LL)$} &
\multicolumn{2}{|c||}{$(\bar RR)(\bar RR)$} &
\multicolumn{2}{|c||}{$(\bar LL)(\bar RR)$}\\ \hline $Q_{ll}$ & $(\bar
l'_p \gamma_\mu l'_r)(\bar l'_s \gamma^\mu l'_t)$ & $Q_{ee}$ & $(\bar
e'_p \gamma_\mu e'_r)(\bar e'_s \gamma^\mu e'_t)$ & $Q_{le}$ & $(\bar
l'_p \gamma_\mu l'_r)(\bar e'_s \gamma^\mu e'_t)$ \\ $Q_{qq}^{(1)}$ &
$(\bar q'_p \gamma_\mu q'_r)(\bar q'_s \gamma^\mu q'_t)$ & $Q_{uu}$ &
$(\bar u'_p \gamma_\mu u'_r)(\bar u'_s \gamma^\mu u'_t)$ & $Q_{lu}$ &
$(\bar l'_p \gamma_\mu l'_r)(\bar u'_s \gamma^\mu u'_t)$
\\ $Q_{qq}^{(3)}$ & $(\bar q'_p \gamma_\mu \tau^I q'_r)(\bar q'_s
\gamma^\mu \tau^I q'_t)$ & $Q_{dd}$ & $(\bar d'_p \gamma_\mu
d'_r)(\bar d'_s \gamma^\mu d'_t)$ & $Q_{ld}$ & $(\bar l'_p \gamma_\mu
l'_r)(\bar d'_s \gamma^\mu d'_t)$ \\ $Q_{lq}^{(1)}$ & $(\bar l'_p
\gamma_\mu l'_r)(\bar q'_s \gamma^\mu q'_t)$ & $Q_{eu}$ & $(\bar e'_p
\gamma_\mu e'_r)(\bar u'_s \gamma^\mu u'_t)$ & $Q_{qe}$ & $(\bar q'_p
\gamma_\mu q'_r)(\bar e'_s \gamma^\mu e'_t)$ \\ $Q_{lq}^{(3)}$ &
$(\bar l'_p \gamma_\mu \tau^I l'_r)(\bar q'_s \gamma^\mu \tau^I q'_t)$
& $Q_{ed}$ & $(\bar e'_p \gamma_\mu e'_r)(\bar d'_s\gamma^\mu d'_t)$ &
$Q_{qu}^{(1)}$ & $(\bar q'_p \gamma_\mu q'_r)(\bar u'_s \gamma^\mu
u'_t)$ \\ && $Q_{ud}^{(1)}$ & $(\bar u'_p \gamma_\mu u'_r)(\bar d'_s
\gamma^\mu d'_t)$ & $Q_{qu}^{(8)}$ & $(\bar q'_p \gamma_\mu
\mathcal{T}^A q'_r)(\bar u'_s \gamma^\mu \mathcal{T}^A u'_t)$ \\ &&
$Q_{ud}^{(8)}$ & $(\bar u'_p \gamma_\mu \mathcal{T}^A u'_r)(\bar d'_s
\gamma^\mu \mathcal{T}^A d'_t)$ & $Q_{qd}^{(1)}$ & $(\bar q'_p
\gamma_\mu q'_r)(\bar d'_s \gamma^\mu d'_t)$ \\ &&&& $Q_{qd}^{(8)}$ &
$(\bar q'_p \gamma_\mu \mathcal{T}^A q'_r)(\bar d'_s \gamma^\mu
\mathcal{T}^A d'_t)$\\ \hline\hline \multicolumn{2}{||c||}{$(\bar
  LR)(\bar RL)$ and $(\bar LR)(\bar LR)$} &
\multicolumn{4}{|c||}{$B$-violating}\\\hline $Q_{ledq}$ & $(\bar
l_p^{'j}e'_r)(\bar d'_s q_t^{'j})$ & $Q_{duq}$ &
\multicolumn{3}{|c||}{$\eps^{\alpha\beta\gamma} \eps_{jk} \left[
    (d^{'\alpha}_p)^T \mathbb{C} u^{'\beta}_r \right]\left[(q^{'\gamma
      j}_s)^T \mathbb{C} l^{'k}_t\right]$}\\ $Q_{quqd}^{(1)}$ & $(\bar
q_p^{'j} u'_r) \eps_{jk} (\bar q_s^{'k} d'_t)$ & $Q_{qqu}$ &
\multicolumn{3}{|c||}{$\eps^{\alpha\beta\gamma} \eps_{jk} \left[
    (q^{'\alpha j}_p)^T \mathbb{C} q^{'\beta k}_r
    \right]\left[(u^{'\gamma}_s)^T \mathbb{C}
    e'_t\right]$}\\ $Q_{quqd}^{(8)}$ & $(\bar q_p^{'j} \mathcal{T}^A
u'_r) \eps_{jk} (\bar q_s^{'k} \mathcal{T}^A d'_t)$ & $Q_{qqq}$ &
\multicolumn{3}{|c||}{$\eps^{\alpha\beta\gamma} \eps_{jn} \eps_{km}
  \left[ (q^{'\alpha j}_p)^T \mathbb{C} q^{'\beta k}_r
    \right]\left[(q^{'\gamma m}_s)^T \mathbb{C}
    l^{'n}_t\right]$}\\ $Q_{lequ}^{(1)}$ & $(\bar l_p^{\prime\, j}
e^\prime_r) \eps_{jk} (\bar q_s^{\prime\, k} u^\prime_t)$ & $Q_{duu}$
& \multicolumn{3}{|c||}{$\eps^{\alpha\beta\gamma} \left[
    (d^{'\alpha}_p)^T \mathbb{C} u^{'\beta}_r
    \right]\left[(u^{'\gamma}_s)^T \mathbb{C}
    e'_t\right]$}\\ $Q_{lequ}^{(3)}$ & $(\bar l_p^{'j} \sigma_{\mu\nu}
e'_r) \eps_{jk} (\bar q_s^{'k} \sigma^{\mu\nu} u'_t)$ & &
\multicolumn{3}{|c||}{}\\ \hline\hline
\end{tabular}
\caption{\sf Four-fermion operators (from \Ref{Grzadkowski:2010es}).
  For brevity we suppress fermion chiral indices
  $L,R$. \label{tab:4ferm}}
\end{table}

\section{Mass eigenstates basis in  SMEFT}
\label{sec:eigen}

As usual, in order to identify physical (and unphysical) degrees of
freedom in the presence of SSB, one needs to diagonalize the resulting
mass matrices for all fields. However, in SMEFT there is an extra
intermediate step involving field rescalings, since SSB also affects
the canonical normalization of the kinetic terms.  In the following
sections we discuss this procedure step by step.

\subsection{Higgs mechanism}
\label{sec:Higgs}

The relevant operator terms contributing to the Higgs potential are
%
\bea
\mathcal{L}_{\mathrm{H}} &= & (D_{\mu} \varphi)^{\dagger} (D^{\mu}
\varphi) + m^{2} (\varphi^{\dagger}\varphi) -
\frac{\lambda}{2}(\varphi^{\dagger}\varphi)^{2} \nonumber \\[2mm]
&+ &C^{\varphi} (\varphi^{\dagger}\varphi)^{3} \ +
\ C^{\varphi\square} (\varphi^{\dagger}\varphi)\square
(\varphi^{\dagger}\varphi) \ + \ C^{\varphi D} (\varphi^{\dagger}
D_{\mu} \varphi)^{*} (\varphi^{\dagger} D^{\mu} \varphi) \;.
\label{eq:LHiggs}
\eea
Minimization of the potential results in a ``corrected" vacuum
expectation value (vev), which reads \cite{Alonso:2013hga},
\bea v = \sqrt{\frac{2m^2}{\lambda}} + \frac{3 m^3
}{\sqrt{2}\lambda^{5/2} } C^\varphi \;.
  \label{eq:vev}
\eea
Notice that in all our expressions and Feynman rules that follow we
use only this vev. As usual, we next expand the Higgs doublet field
around the vacuum,
\begin{equation}
\varphi = \left ( \begin{array}{c} \Phi^+ \\ \frac{1}{\sqrt{2}} (v + H
  + i \Phi^0) \end{array} \right ) \;.
\end{equation}
%
The Lagrangian bilinear terms of the scalar fields are then given by,
\bea
\mathcal{L}^{\mathrm {Bilinear}}_{H} &=& {1\over 2}\left( 1+{1\over 2}
C^{\varphi D}v^2 - 2 {C^{\varphi\Box}} v^2\right) (\partial_\mu H)^2
+\left({1\over 2}m^2 - {3\over 4}\lambda v^2+{15\over 8} v^4 C^\varphi
\right) H^2 \nonumber\\[2mm]
&+& {1\over 2}\left(1+{1\over 2}C^{\varphi D}v^2\right)(\partial_\mu
\Phi^{0})^2 + (\partial_\mu \Phi^{-})(\partial^\mu \Phi^{+}).
\eea
By rescaling the fields as
\begin{equation}
h = Z_h\, H\;, \qquad G^0 = Z_{G^0}\, \Phi^0\;, \qquad G^\pm \equiv
\Phi^\pm \;,
\end{equation}
with the constant factors
\bea
Z_{h} &\equiv& 1+{1\over 4} C^{\varphi D}v^2-C^{\varphi\Box} v^2 \;,
\label{eq:Zh} \\
Z_{G^0} &\equiv& 1+{1\over 4}C^{\varphi D}v^2\;,
\label{eq:ZphiD}
\eea
one obtains the physical Higgs field $h$ and Goldstone fields
$G^0,G^\pm$ with canonically normalized kinetic terms. The tree-level
squared mass of the normalized Higgs field $h$ now reads,
\bea
M_h^{2} &=& 2 m^2\left[1 - \frac{m^2}{\lambda^{2}} \left(3\,
  C^{\varphi} - 4\, \lambda \, C^{\varphi\square} + \lambda \,
  C^{\varphi D} \right)\right]\nn
&=& \lambda v^2 - (3\, C^{\varphi} - 2\, \lambda\, C^{\varphi\square}
+ \frac{\lambda}{2}\, C^{\varphi D} ) v^4 \;.
\eea
%

\subsection{The gauge sector}
\label{sec:EW}

The Lagrangian terms which are relevant for gauge boson propagators
read,
\begin{align}
\mathcal{L}_{\mathrm{EW}} &= -\frac{1}{4} W_{\mu\nu}^I W^{I\mu\nu} -
\frac{1}{4} B_{\mu\nu} B^{\mu\nu} +(D_{\mu} \varphi)^{\dagger}
(D^{\mu} \varphi) \nonumber \\[2mm]
&+ C^{\varphi W} (\varphi^\dagger \varphi) W_{\mu\nu}^I W^{I\mu\nu} +
C^{\varphi B} (\varphi^\dagger \varphi) B_{\mu\nu} B^{\mu\nu} +
C^{\varphi W B} (\varphi^\dagger \tau^I \varphi) W_{\mu\nu}^I
B^{\mu\nu} \nonumber \\[2mm]
&+ C^{\varphi D} (\varphi^{\dagger} D_{\mu} \varphi)^{*}
(\varphi^{\dagger} D^{\mu} \varphi)\;, \label{Lag:EW} \\[3mm]
\mathcal{L}_{\mathrm{QCD}} &= -\frac{1}{4} G_{\mu\nu}^A G^{A\mu\nu} +
C^{\varphi G} (\varphi^\dagger \varphi) G_{\mu\nu}^A G^{A\mu\nu} \;,
\label{eq:lgauge}
\end{align}
where $\tau^I$ are the Pauli matrices. Other, potentially relevant
operators of the theory, containing $\widetilde{B}_{\mu\nu}$,
$\widetilde{W}_{\mu\nu}^I$ and $\widetilde{G}_{\mu\nu}^A$ influence
only CP-violating vertices. Their bilinear terms are total derivatives
and do not affect propagators. Therefore, we neglect them in our
discussion here.

To simplify the above expressions, it is convenient to introduce
``barred'' fields and couplings, such as
\bea
\begin{array}{ll}
\bar W_\mu^I \equiv Z_g W_\mu^I , \qquad & \bar g\, \equiv Z_{g}^{-1}
g ~, \nn
\bar B_\mu\equiv Z_{g'} B_\mu ~, \qquad & \bar g' \equiv Z_{g'}^{-1}
g' ,\nn
\bar G_\mu^A \equiv Z_{g_s} G_\mu^A , \qquad & \bar g_s \equiv
Z_{g_s}^{-1} g_s\, ,
\end{array}
\label{Zg-norm} 
\eea
where for our constant, field and coupling rescalings, we choose
\bea
Z_{g} &\equiv& 1-C^{\varphi W}v^2\;,\nn Z_{g'} &\equiv& 1-C^{\varphi
  B}v^2 \;, \label{eq:zgdef} \\ Z_{g_s} &\equiv& 1-C^{\varphi
  G}v^2\;.\nonumber
\eea
We note that such transformations do not violate gauge invariance.
They preserve the form of the covariant derivative which now reads,
\begin{align}
D_\mu= \bar{D}_\mu = \partial_\mu + i \bar{g}\bar B_\mu Y+i
\bar{g}\bar W_\mu^I T^I + i\bar g_s \bar G_\mu^A \mathcal{T}^A \;,
\end{align} 
while the field strength tensors rescale the same way as their
respective fields.  The particular choice of~\eq{eq:zgdef} renders the
kinetic terms for the electroweak fields canonical, with an exception
of the mixed $Q_{\varphi W B}$ operator in \eq{Lag:EW}.  Furthermore,
the last redefinition of~\eq{eq:zgdef} is sufficient to define
massless physical, canonically normalized gluon fields, as
\bea
g_\mu^A \equiv \bar G_\mu^A \;.
\eea
In terms of ``barred'' electroweak gauge bosons, $\bar B_\mu$ and
$\bar W_\mu$, the bilinear part of the Lagrangian reads,
\bea
\mathcal{L}_{EW}^{\mathrm{Bilinear}} &= & -{1\over 4} (\bar
W^1_{\mu\nu}\bar W^{1\mu\nu}+\bar W^2_{\mu\nu}\bar W^{2\mu\nu} )
-{1\over 4} \left(\begin{array}{c} \bar W^3_{\mu\nu} \\ \bar
  B_{\mu\nu} \\
\end{array}\right)^\top\left(\begin{array}{cc}
 1 & \epsilon \\ \epsilon & 1
\end{array}\right) \left(\begin{array}{c}
\bar W^{3\mu\nu} \\ \bar B^{\mu\nu} \\
\end{array}\right)  \nn
&& + \f{\bar g^2 v^2}{8}(\bar W^1_{\mu}\bar W^{1\mu}+\bar
W^2_{\mu}\bar W^{2\mu}) \nonumber \\ &&+ {v^2\over 8} Z_{G^0}^2
\left(\begin{array}{c} \bar W^3_{\mu} \\ \bar B_{\mu} \\
\end{array}\right)^\top
\left(\begin{array}{rr} \bar g^2 & -\bar g\bar g' \\ -\bar g\bar g' &
  \bar g^{'2}
\end{array}\right)
\left(\begin{array}{c} \bar W^{3\mu} \\ \bar B^{\mu} \\
\end{array}\right) \;,  \label{L-WBmass}
\eea
where we have defined,
\begin{align}
 \epsilon \equiv C^{\varphi WB} ~v^2 \;.
 \label{eq:eps}
\end{align}
From \eq{L-WBmass} one identifies immediately the physical charged
gauge bosons ${W}_\mu^\pm$, as
\begin{equation}
{W}_\mu^\pm = {1\over \sqrt{2}}(\bar W_\mu ^1 \mp i \bar W_\mu
^2)\;, \label{Wpm-rot}
\end{equation}
with the mass
\begin{equation}
M_W = \frac{1}{2}\, \bar g \, v \;.
\label{eq:mw}
\end{equation}
The neutral gauge boson mass basis is obtained through the congruent
matrix transformation~\cite{Horn}, producing simultaneously canonical
kinetic terms and diagonal masses. It reads,
\bea
\left ( \begin{array}{c} \bar{W}^3_\mu \\ \bar{B}_\mu \end{array}
\right ) &=\ \mathbb{X} \, \left (\begin{array}{c} {Z}_\mu
  \\ {A}_\mu \end{array} \right ) \;,
\label{eq:rot}
\eea
with the matrix $\mathbb{X}$ taking the form,
\bea
\mathbb{X} = \left(%
\begin{array}{cc}
 1&-{\epsilon\over 2} \\ -{\epsilon\over 2} &1
 \end{array}%
\right) \left(%
\begin{array}{cc}
 \cos{\bar{\theta}}& \sin{\bar{\theta}} \\ -\sin{\bar{\theta}} &
 \cos{\bar{\theta}}
 \end{array}%
\right)\;.
\label{eq:X}
\eea
Straightforward calculation leads to a mixing
angle~\cite{Grinstein:1991cd,Alonso:2013hga}
\bea
\tan \bar{\theta} = \frac{\bar g'}{\bar g} + \frac{\epsilon}{2}\left(1
- \frac{\bar g^{'2}}{ \bar g^{2} }\right) \;,\label{eq:theta}
\eea
whereas for gauge boson masses we obtain
\bea
M_Z &=& \frac{1}{2} \, \sqrt{\bar g^2 + \bar g^{'2}} \, v \, \left(1 +
\frac{\epsilon \bar g \bar g'}{\bar g^2 + \bar g^{'2}}\right) \,
Z_{G^0}\;, \nn
M_A &=& 0\;.
\label{eq:zmass}
\eea
One can easily verify that the photon remains massless from the
vanishing determinant of the mass matrix in \eq{L-WBmass}. Note also
that the $\mathbb{X}$ transformation affects the trace of this matrix,
thus producing the $\epsilon$-dependence for $M_Z$.

\subsection{Gauge-Goldstone mixing}
\label{sec:goldstone}

The operators relevant for Goldstone bosons kinetic terms give also
rise to Goldstone-gauge boson mixing. They read,
\bea
\mathcal{L}_{H} \supset \ (\bar D_{\mu} \varphi)^{\dagger} (\bar
D^{\mu} \varphi)\ +\ C^{\varphi D} (\varphi^{\dagger} \bar D_{\mu}
\varphi)^{*} (\varphi^{\dagger}\bar D^{\mu} \varphi) \;,
\eea
which, in the presence of SSB, generate the ``unwanted" terms
\bea
\mathcal{L}_{G-EW } &= & - \ i {\bar g v\over 2\sqrt{2}}\, \bar
W^1_\mu\Big(\partial^\mu \Phi^+ - \partial^\mu \Phi^- \Big)+{\bar g
  v\over 2\sqrt{2}}\, \bar W^2_\mu\Big(\partial^\mu \Phi^+
+\ \partial^\mu \Phi^- \Big)\nonumber\\
&& - \ {\bar gv\over 2} Z_{G_0}^2\, \bar W^3_{\mu} \,
\partial^\mu\Phi^0 + \ {\bar g' v\over 2} Z_{G_0}^2\, \bar B_{\mu}\,
\partial^\mu\Phi^0 .\label{L-dphiWB}
\eea
After expressing $\mathcal{L}_{G-EW}$ in terms of the physical gauge
bosons and Goldstone bosons, one arrives to the familiar expression,
\bea
\mathcal{L}_{G-EW } &=& i M_W( W^+_\mu \partial^\mu G^- - W^-_\mu
\partial^\mu G^+)
- M_Z\, Z_\mu\, \partial^\mu G^0 \;.\label{eq:GEW}
\eea
Thus, in mass basis all Wilson coefficients in the bilinear
gauge-Goldstone mixing have been absorbed in the definitions of fields
and masses.  As we discuss in Section~\ref{sec:gaugefix}, such a
property essentially allows to adopt the standard $R_\xi$-gauge fixing
also for SMEFT loop calculations.

\subsection{Fermion sector}
\label{sec:fmass}

The operators relevant to fermion masses are
\begin{eqnarray}
\mathcal{L}_{f} &=& i (\bar{l}^\prime_L\, \slashed{\bar{D}}\,
l_L^\prime + \bar{e}_R^\prime\, \slashed{\bar{D}} \, e_R^\prime +
\bar{q}_L^\prime\, \slashed{\bar{D}} \, q^\prime_L + \bar{u}_R^\prime
\, \slashed{\bar{D}}\, u_R^\prime + \bar{d}_R^\prime \,
\slashed{\bar{D}} \, d_R^\prime ) \nonumber \\[2mm]
&-& (\bar{l}^\prime_L\, {\Gamma_e} \, {e}_R^\prime\, \varphi +
\bar{q}^\prime_L\, {\Gamma_u} \, {u}_R^\prime \, \widetilde{\varphi} +
\bar{q}^\prime_L\, {\Gamma_d} \, {d}_R^\prime \, {\varphi} +
\mathrm{H.c.})
 \label{eq:LY} \nonumber  \\[2mm]
&+& \left [ (\varphi^\dagger \varphi) \, ( \bar{l}_L^\prime \,
   C^{'e\varphi} \, e_R^\prime \, \varphi ) + (\varphi^\dagger
   \varphi) \, ( \bar{q}_L^\prime \, C^{'u\varphi} \, u_R^\prime \,
   \widetilde{\varphi}) + (\varphi^\dagger \varphi) \, (
   \bar{q}_L^\prime \, C^{'d\varphi} \, d_R^\prime \, \varphi ) +
   \mathrm{H.c.} \right ] \nonumber\\[2mm]
&+& \left [ C^{'\nu\nu} \: (\widetilde{\varphi}^\dagger\:
   l_L^\prime)^T \: \mathbb{C} \: (\widetilde{\varphi}^\dagger \:
   l_L^\prime) + \mathrm{H.c.}\right ] \;, \label{eq:WO}
\end{eqnarray}
where $\Gamma_{e,u,d}$ and $C^{'e\varphi}, C^{'u\varphi},
C^{'d\varphi}$ are general complex $3\times 3$ matrices, $C^{'\nu\nu}$
is a symmetric complex $3\times 3$ matrix and primed fields denote the
fields in the interaction (gauge) basis (group and generation indices
are suppressed).

The fermion kinetic terms remain unaffected by SSB, while the mass
terms read
\begin{eqnarray}
\mathcal{L}_{\mathrm{mass}} = - \frac{1}{2}\, \nu_L^{\prime\, T}\,
\mathbb{C}\, M^\prime_\nu \, \nu_L^\prime - \bar{e}_L^\prime \,
M^\prime_e \, e_R^\prime - \bar{u}_L^\prime \, M^\prime_u \,
u_R^\prime - \bar{d}_L^\prime \, M^\prime_d \, d_R^\prime +
\mathrm{H.c.} \;,
\end{eqnarray}
with the $3\times 3$ mass matrices equal to
\begin{equation}
\begin{array}{cc}
M^\prime_\nu = - v^2 C^{'\nu\nu}\;, &
M^\prime_e = \frac{v}{\sqrt{2}}\, \left (\Gamma_e - C^{'e\varphi}
\frac{v^2}{2}\right ),\; \\[2mm]
M^\prime_u = \frac{v}{\sqrt{2}}\, \left (\Gamma_u - C^{'u\varphi}
\frac{v^2}{2}\right ),\; &
M^\prime_d = \frac{v}{\sqrt{2}}\, \left (\Gamma_d - C^{'d\varphi}
\frac{v^2}{2}\right )\;.
\end{array}
\label{eq:modyuk}
\end{equation}
To diagonalize lepton and quark masses we rotate the fermion fields by
the unitary matrices,
\begin{equation}
\psi_X^\prime = U_{\psi_X} \: \psi_X\;,
\end{equation}
with $\psi=\nu,e,u,d$, $X=L,R$ and the ``unprimed" symbols denoting
the mass eigenstates fields.  Then, the singular value decomposition
for charged fermion mass matrices results in
\begin{eqnarray}
U_{e_L}^\dagger \, M^\prime_e \, U_{e_R} =& {M}_e = \mathrm{diag}(m_e,
m_\mu, m_\tau) \;, \nonumber\\[2mm]
U_{u_L}^\dagger \, M^\prime_u \, U_{u_R} =& {M}_u = \mathrm{diag}(m_u,
m_c, m_t) \;, \label{eq:uu} \\[2mm]
U_{d_L}^\dagger \, M^\prime_d \, U_{d_R} =& {M}_d = \mathrm{diag}(m_d,
m_s, m_b) \;, \nonumber
\end{eqnarray}
while the diagonal neutrino mass matrix is obtained through
\begin{equation}
 U^T_{\nu_L} \: M^\prime_{{\nu}} \: U_{\nu_L} ~=~ {M}_{\nu} =
 \mathrm{diag}(m_{\nu_1},m_{\nu_2}, m_{\nu_3}) \;,
\end{equation}
with all fermion masses now being real and non-negative.

It is important to stress that in the absence of fermion
  SM-singlets in the SMEFT spectrum, neutrinos can never form massive
  Dirac particles. In the general case of $L$-flavour and number
  violation, through the Weinberg operator of~\eq{qnunu}, neutrinos
  are massive Majorana spinors whereas under the assumption of
  $L$-conservation they can also be regarded just as massless Weyl
  spinors. However, if dimension-six SMEFT is intended to describe the
  observable neutrino effects then the choice of massive Majorana
  neutrinos becomes mandatory.  In the massless limit the Majorana
  formalism becomes physically equivalent to the Weyl approach.
  Therefore, considering neutrinos as Majorana particles in our
  Feynman rules covers all possible consistent choices.  Such an
  approach requires special set of rules for neutrino propagators,
  vertices and diagram combinatorics (the latter being the same as for
  the real-scalar fields). We follow here the treatment proposed by
  Denner \textit{ et al.,}\cite{Denner:1992vza,Denner:1992me}. More
  details are given in Appendix \ref{app:fermion}.

We note also that in the Majorana case \textit{all} interactions
involving neutrinos are automatically lepton number violating.
Obviously, the $L$-violating effects are controlled by neutrino
masses, giving typically small or negligible corrections to the
amplitudes.

\begin{table}
  \begin{tabular}{|l|l|}
    \hline
$C^{e\varphi} = U_{e_L}^\dagger C^{'e\varphi} U_{e_R}$ &
    $(C^{ll})_{f_1 f_2 f_3 f_4} = (U_{e_L})_{g_2 f_2} (U_{e_L})_{g_4
      f_4} (U_{e_L})_{g_1 f_1}^* (U_{e_L})_{g_3 f_3}^* (C^{'ll})_{g_1
      g_2 g_3 g_4} $ \\[1.5mm]
$C^{d\varphi} = U_{d_L}^\dagger C^{'d\varphi} U_{d_R}$ &
    $(C^{ee})_{f_1 f_2 f_3 f_4} = (U_{e_R})_{g_2 f_2} (U_{e_R})_{g_4
      f_4} (U_{e_R})_{g_1 f_1}^* (U_{e_R})_{g_3 f_3}^* (C^{'ee})_{g_1
      g_2 g_3 g_4} $ \\[1.5mm]
$C^{u\varphi} = U_{u_L}^\dagger C^{'u\varphi} U_{u_R}$ &
    $(C^{le})_{f_1 f_2 f_3 f_4} = (U_{e_L})_{g_2 f_2} (U_{e_R})_{g_4
      f_4} (U_{e_L})_{g_1 f_1}^* (U_{e_R})_{g_3 f_3}^* (C^{'le})_{g_1
      g_2 g_3 g_4} $ \\[1.5mm]
$C^{eW} = U_{e_L}^\dagger C^{'eW} U_{e_R}$ & $(C^{qq (1)})_{f_1 f_2
      f_3 f_4} = (U_{d_L})_{g_2 f_2} (U_{d_L})_{g_4 f_4}(U_{d_L})_{g_1
      f_1}^* (U_{d_L})_{g_3 f_3}^* (C^{'qq (1)})_{g_1 g_2 g_3 g_4} $
    \\[1.5mm]
$C^{eB} = U_{e_L}^\dagger C^{'eB} U_{e_R}$ & $(C^{qq (3)})_{f_1 f_2
      f_3 f_4} = (U_{d_L})_{g_2 f_2} (U_{d_L})_{g_4 f_4}
    (U_{d_L})_{g_1 f_1}^* (U_{d_L})_{g_3 f_3}^* (C^{'qq (3)})_{g_1
      g_2 g_3 g_4} $ \\[1.5mm]
$C^{dG} = U_{d_L}^\dagger C^{'dG} U_{d_R}$ & $(C^{dd})_{f_1 f_2 f_3
      f_4} = (U_{d_R})_{g_2 f_2} (U_{d_R})_{g_4 f_4} (U_{d_R})_{g_1
      f_1}^* (U_{d_R})_{g_3 f_3}^* (C^{'dd})_{g_1 g_2 g_3 g_4} $
    \\[1.5mm]
$C^{dW} = U_{d_L}^\dagger C^{'dW} U_{d_R}$ & $(C^{uu})_{f_1 f_2 f_3
      f_4} = (U_{u_R})_{g_2 f_2} (U_{u_R})_{g_4 f_4} (U_{u_R})_{g_1
      f_1}^* (U_{u_R})_{g_3 f_3}^* (C^{'uu})_{g_1 g_2 g_3 g_4} $
    \\[1.5mm]
$C^{dB} = U_{d_L}^\dagger C^{'dB} U_{d_R}$ & $(C^{ud (1)})_{f_1 f_2
      f_3 f_4} = (U_{u_R})_{g_2 f_2} (U_{d_R})_{g_4 f_4}
    (U_{u_R})_{g_1 f_1}^* (U_{d_R})_{g_3 f_3}^* (C^{'ud (1)})_{g_1
      g_2 g_3 g_4} $ \\[1.5mm]
$C^{uG} = U_{u_L}^\dagger C^{'uG} U_{u_R}$ & $(C^{ud (8)})_{f_1 f_2
      f_3 f_4} = (U_{u_R})_{g_2 f_2} (U_{d_R})_{g_4 f_4}
    (U_{u_R})_{g_1 f_1}^* (U_{d_R})_{g_3 f_3}^* (C^{'ud (8)})_{g_1 g_2
      g_3 g_4} $ \\[1.5mm]
$C^{uW} = U_{u_L}^\dagger C^{'uW} U_{u_R}$ & $(C^{qu (1)})_{f_1 f_2
      f_3 f_4} = (U_{d_L})_{g_2 f_2} (U_{u_R})_{g_4 f_4}
    (U_{d_L})_{g_1 f_1}^* (U_{u_R})_{g_3 f_3}^* (C^{'qu (1)})_{g_1
      g_2 g_3 g_4} $ \\[1.5mm]
$C^{uB} = U_{u_L}^\dagger C^{'uB} U_{u_R}$ & $(C^{qu (8)})_{f_1 f_2
      f_3 f_4} = (U_{d_L})_{g_2 f_2} (U_{u_R})_{g_4 f_4}
    (U_{d_L})_{g_1 f_1}^* (U_{u_R})_{g_3 f_3}^* (C^{'qu (8)})_{g_1
      g_2 g_3 g_4} $ \\[1.5mm]
$C^{\varphi l (1)} = U_{e_L}^\dagger C^{'\varphi l (1)} U_{e_L}$ &
    $(C^{qd (1)})_{f_1 f_2 f_3 f_4} = (U_{d_L})_{g_2 f_2}
    (U_{d_R})_{g_4 f_4} (U_{d_L})_{g_1 f_1}^* (U_{d_R})_{g_3 f_3}^*
    (C^{'qd (1)})_{g_1 g_2 g_3 g_4} $ \\[1.5mm]
$C^{\varphi l (3)} = U_{e_L}^\dagger C^{'\varphi l (3)} U_{e_L}$ &
    $(C^{qd (8)})_{f_1 f_2 f_3 f_4} = (U_{d_L})_{g_2 f_2}
    (U_{d_R})_{g_4 f_4} (U_{d_L})_{g_1 f_1}^* (U_{d_R})_{g_3 f_3}^*
    (C^{'qd (8)})_{g_1 g_2 g_3 g_4} $ \\[1.5mm]
$C^{\varphi e} = U_{e_R}^\dagger C^{'\varphi e} U_{e_R}$ & $(C^{quqd
      (1)})_{f_1 f_2 f_3 f_4} = (U_{u_R})_{g_2 f_2} (U_{d_R})_{g_4
      f_4} (U_{d_L})_{g_1 f_1}^* (U_{d_L})_{g_3 f_3}^* (C^{'quqd
      (1)})_{g_1 g_2 g_3 g_4} $ \\[1.5mm]
$C^{\varphi q (1)} = U_{d_L}^\dagger C^{'\varphi q (1)} U_{d_L}$ &
    $(C^{quqd (8)})_{f_1 f_2 f_3 f_4} = (U_{u_R})_{g_2 f_2}
    (U_{d_R})_{g_4 f_4} (U_{d_L})_{g_1 f_1}^* (U_{d_L})_{g_3 f_3}^*
    (C^{'quqd (8)})_{g_1 g_2 g_3 g_4} $ \\[1.5mm]
$C^{\varphi q (3)} = U_{d_L}^\dagger C^{'\varphi q (3)} U_{d_L}$ &
    $(C^{lq (1)})_{f_1 f_2 f_3 f_4} = (U_{e_L})_{g_2 f_2}
    (U_{d_L})_{g_4 f_4}(U_{e_L})_{g_1 f_1}^* (U_{d_L})_{g_3 f_3}^*
    (C^{'lq (1)})_{g_1 g_2 g_3 g_4} $ \\[1.5mm]
$C^{\varphi d} = U_{d_R}^\dagger C^{'\varphi d} U_{d_R}$ & $(C^{lq
      (3)})_{f_1 f_2 f_3 f_4} = (U_{e_L})_{g_2 f_2} (U_{d_L})_{g_4
      f_4} (U_{e_L})_{g_1 f_1}^* (U_{d_L})_{g_3 f_3}^* (C^{'lq
      (3)})_{g_1 g_2 g_3 g_4} $ \\[1.5mm]
$C^{\varphi u} = U_{u_R}^\dagger C^{'\varphi u} U_{u_R}$ &
    $(C^{ld})_{f_1 f_2 f_3 f_4} = (U_{e_L})_{g_2 f_2} (U_{d_R})_{g_4
      f_4} (U_{e_L})_{g_1 f_1}^* (U_{d_R})_{g_3 f_3}^* (C^{'ld})_{g_1
      g_2 g_3 g_4} $ \\[1.5mm]
$C^{\varphi ud} = U_{u_R}^\dagger C^{'\varphi ud} U_{d_R}$ &
    $(C^{lu})_{f_1 f_2 f_3 f_4} = (U_{e_L})_{g_2 f_2} (U_{u_R})_{g_4
      f_4} (U_{e_L})_{g_1 f_1}^* (U_{u_R})_{g_3 f_3}^* (C^{'lu})_{g_1
      g_2 g_3 g_4} $ \\[1.5mm]
$C^{\nu\nu} = U_{\nu_L}^\top C^{'\nu\nu} U_{\nu_L}$ & $(C^{qe})_{f_1 f_2
      f_3 f_4} = (U_{d_L})_{g_2 f_2} (U_{e_R})_{g_4 f_4}
    (U_{d_L})_{g_1 f_1}^* (U_{e_R})_{g_3 f_3}^* (C^{'qe})_{g_1 g_2 g_3
      g_4} $ \\[1.5mm]
& $(C^{ed})_{f_1 f_2 f_3 f_4} = (U_{e_R})_{g_2 f_2} (U_{d_R})_{g_4
      f_4} (U_{e_R})_{g_1 f_1}^* (U_{d_R})_{g_3 f_3}^* (C^{'ed})_{g_1
      g_2 g_3 g_4} $ \\[1.5mm]
& $(C^{eu})_{f_1 f_2 f_3 f_4} = (U_{e_R})_{g_2 f_2} (U_{u_R})_{g_4
      f_4} (U_{e_R})_{g_1 f_1}^* (U_{u_R})_{g_3 f_3}^* (C^{'eu})_{g_1
      g_2 g_3 g_4} $ \\[1.5mm]
& $(C^{ledq})_{f_1 f_2 f_3 f_4} = (U_{e_R})_{g_2 f_2} (U_{d_L})_{g_4
      f_4} (U_{e_L})_{g_1 f_1}^* (U_{d_R})_{g_3 f_3}^*
    (C^{'ledq})_{g_1 g_2 g_3 g_4} $ \\[1.5mm]
& $(C^{lequ (1)})_{f_1 f_2 f_3 f_4} = (U_{e_R})_{g_2 f_2}
    (U_{u_R})_{g_4 f_4} (U_{e_L})_{g_1 f_1}^* (U_{d_L})_{g_3 f_3}^*
    (C^{'lequ (1)})_{g_1 g_2 g_3 g_4} $ \\[1.5mm]
& $(C^{lequ (3)})_{f_1 f_2 f_3 f_4} = (U_{e_R})_{g_2 f_2}
    (U_{u_R})_{g_4 f_4} (U_{e_L})_{g_1 f_1}^* (U_{d_L})_{g_3 f_3}^*
    (C^{'lequ (3)})_{g_1 g_2 g_3 g_4} $ \\[1.5mm]
& $(C^{duq})_{f_1 f_2 f_3 f_4} = (U_{u_R})_{g_2 f_2} (U_{e_L})_{g_4
      f_4} (U_{d_R})_{g_1 f_1} (U_{d_L})_{g_3 f_3} (C^{'duq})_{g_1 g_2
      g_3 g_4} $ \\[1.5mm]
& $(C^{qqu})_{f_1 f_2 f_3 f_4} = (U_{d_L})_{g_2 f_2}] (U_{e_R})_{g_4
        f_4} (U_{d_L})_{g_1 f_1} (U_{u_R})_{g_3 f_3} (C^{'qqu})_{g_1
        g_2 g_3 g_4} $ \\[1.5mm]
& $(C^{qqq})_{f_1 f_2 f_3 f_4} = (U_{d_L})_{g_2 f_2} (U_{e_L})_{g_4
        f_4} (U_{d_L})_{g_1 f_1} (U_{d_L})_{g_3 f_3} (C^{'qqq})_{g_1
        g_2 g_3 g_4} $ \\[1.5mm]
& $(C^{duu})_{f_1 f_2 f_3 f_4} = (U_{u_R})_{g_2 f_2} (U_{e_R})_{g_4
        f_4} (U_{d_R})_{g_1 f_1} (U_{u_R})_{g_3 f_3} (C^{'duu})_{g_1
        g_2 g_3 g_4} $ \\[1.5mm] \hline
\end{tabular}
\caption{\sf Definitions for the Wilson coefficients of operators involving fermions, in mass basis. We suppress the flavor indices for the two-fermion operators as the contraction is non-ambiguous here and we assume summation over repeating indices. \label{tab:redef}}
\end{table}

\section{Corrections to the SM couplings}
\label{sec:SMcoup}

Corrections to the interactions described by the dimension-4 SM
Lagrangian can come either as genuine new vertices generated by higher
order operators, or from the dimension-4 vertices modified by the
shifts in the fields and parameters necessary to express them in the
mass eigenstates basis. In this section we discuss the second class of
(``oblique'') corrections.

In terms of physical gauge bosons, the electroweak part of the
covariant derivative (its QCD part parametrized in terms of $\bar
g_s$-coupling is unchanged compared to the SM), reads
\begin{align}
\bar{D}_\mu^{EW} =& \ \partial_\mu  +  i \frac{\bar{g}}{\sqrt{2}} \: ( T^+
W_\mu^+ + T^- W_\mu^- ) \nonumber \\[2mm]
+& i (\bar{g} \mathbb{X}_{11}\  T^3  +
\bar{g}^\prime  \mathbb{X}_{21} \  Y )  \: Z_\mu + i (\bar{g} 
\mathbb{X}_{12}\  T^3 + \bar{g}^\prime  \mathbb{X}_{22}\  Y ) \: A_\mu \;.
\end{align} 
The pattern of electroweak symmetry breaking results in a conserved
electric charge, identified through the standard relation
$Q=T_3+Y$. The electromagnetic gauge invariance of the broken theory
manifests through the ``corrected" electroweak unification condition,
\begin{equation}
\bar{e} = \bar{g}^\prime \: \mathbb{X}_{22}  =  \bar{g}\: \mathbb{X}_{12}\;,
\label{eq:uni}
\end{equation}
which couples the photon only to the electric charge while keeping it
massless.  Using \eq{eq:uni} and the property $\det{\mathbb{X}}=1$ one
can always express the covariant derivative in the familiar form,
\begin{equation}
\bar{D}_\mu^{EW} = \partial_\mu + i \frac{\bar{g}}{\sqrt{2}} \: ( T^+
W_\mu^+ + T^- W_\mu^- ) + i \bar{g}_{Z} \: (T^3 - \sin^2\bar
\theta \: Q) Z_\mu+ i \bar{e} Q A_\mu \;,
\label{eq:covder}
\end{equation}
where the modified couplings now read,
\bea
\bar e &=&
\frac{\bar g \bar g'}{\sqrt{\bar g^2 + \bar g^{'2}}}\left(1 -
\frac{\epsilon \bar g \bar g^{'}}{\bar g^2 + \bar g^{'2}}\right)\;,
\nn
\bar g_Z &=&
\sqrt{\bar g^2 + \bar g^{'2}} \left(1 + \frac{\epsilon \bar g \bar
  g^{'}}{\bar g^2 + \bar g^{'2}} \right)\;.
\eea

In summary, after redefinitions of fields and couplings in mass basis,
corrections to gauge interactions originating from the shift in the
gauge and Higgs sector parameters depend only on two additional Wilson
coefficients: $C^{\varphi WB}$, responsible for the mixing of
electroweak gauge boson kinetic terms, and, $C^{\varphi D}$ appearing
through the physical $Z^0$-boson mass (see \eq{eq:zmass}).
Furthermore, the $C^{\varphi D}$ operator breaks the custodial
invariance as this is described by the anomalous value of the $\rho$
parameter,
\begin{equation}
\rho = \frac{|J_{C.C}|^2}{|J_{N.C.}|^2} = \frac{\bar{g}^2\:
  M_{{Z}}^2}{\bar{g}_{{Z}}^2 M_W^2} = 1 + {1\over 2}C^{\varphi D} v^2
\;.
\end{equation}  
As it is well known, this is strongly constrained by precision EW
experiments, at the level of 0.1\%~\cite{PDG2016}.  Consequently,
sizable ``oblique'' corrections in the gauge sector could potentially
arise only from the gauge boson kinetic mixing $\epsilon$ defined in
\eq{eq:eps}.

Another set of ``oblique'' corrections originates in the flavor sector
of SMEFT after diagonalization of the fermion mass matrices [see
  section~\ref{sec:fmass}]. In SM, the flavor rotations obtained from fermion mass diagonalization appear explicitly only in the charged quark and lepton currents  in the specific combinations,
\bea
K\equiv U_{u_L}^\dagger U_{d_L} ~ ,~~   U \equiv U_{e_L}^\dagger U_{\nu_L}~, \label{eq:ckmpmns}
\eea 
and are identified as the CKM~\cite{Kobayashi:1973fv} and
PMNS~\cite{Pontecorvo:1957qd,Maki:1962mu} mixing matrices,
respectively. We follow the same definition in SMEFT\footnote{Alternatively one can follow a non-unitary definition for CKM and PMNS as in previous versions of this manuscript and Ref.~\cite{Aebischer:2015fzz}. The unitary definition adopted here allows for easier manipulations in practical calculations and displays the physical parameters of the flavor sector in a more transparent way.} for obviously different flavor rotations $U_{\psi_X}$ since the corresponding mass matrices are modified, as seen in \eq{eq:modyuk}. Nevertheless, contrary to SM, the \textit{unitary} matrices $K,U$ of SMEFT are not exclusively responsible for the $W$-fermion-fermion vector couplings of quarks and leptons.  This can be seen from the relevant part
of the Lagrangian,
\begin{eqnarray}
\mathcal{L}_{c.c.}  = &-& \frac{\bar{g}}{\sqrt{2}} \; W_\mu^+ \:
\bar{u}_p \: \gamma^\mu \: \left\{ \left[ U_{u_L}^\dagger (\mathbb{1}
  + v^2 C^{'\varphi q (3)}) U_{d_L}\right]_{pr} P_L +
\left(\dfrac{v^2}{2} U_{u_R}^\dagger C^{'\varphi u d}
U_{d_R}\right)_{pr} P_R \right\} \, d_r \nonumber \\[2mm]
&-& \frac{\bar{g}}{\sqrt{2}} \; W_\mu^+ \: \bar{\nu}_p \: \gamma^\mu
\: \left[ U_{e_L}^\dagger ( \mathbb{1} + v^2 C^{'\varphi l (3)})
U_{\nu_L}\right]^\dagger_{pr} \: P_L \: e_r + \mathrm{H.c.}\, ,
\end{eqnarray}  
by substituting \eq{eq:ckmpmns} and performing the parameter redefinitions of Table~\ref{tab:redef} to obtain,
\begin{eqnarray}
\mathcal{L}_{c.c.}  = &-& \frac{\bar{g}}{\sqrt{2}} \; W_\mu^+ \:
\bar{u}_p \: \gamma^\mu \: \left\{ \left[ K\,(\mathbb{1} + v^2 C^{\varphi q (3)})\right]_{pr} P_L +
\dfrac{v^2}{2}  C^{\varphi u d}_{pr} P_R \right\} \, d_r \nonumber \\[2mm]
&-& \frac{\bar{g}}{\sqrt{2}} \; W_\mu^+ \: \bar{\nu}_p \: \gamma^\mu
\: \left[(\mathbb{1} + v^2 C^{\varphi l (3)})\, U\right]^\dagger_{pr} \: P_L \: e_r + \mathrm{H.c.}
\end{eqnarray} 
It should be clear in this form that the Wilson coefficients not only generate novel right-handed interactions but also violate unitarity in the left-handed couplings. 

In what follows we redefine the Wilson coefficients of operators involving fermions by absorbing into them the flavor rotations from gauge to mass basis. In this way we
are able to express the mass basis Lagrangian entirely in terms of the
``unprimed'' fields, Wilson coefficients and the unitary $K$-, and $U$-mixing
matrices. In some cases the redefinitions are not unique, as in the
operators involving left fermion $SU(2)$ doublets one can adsorb into
the Wilson coefficient either the rotation matrix of the lower or
upper constituent of the doublet. We choose it always to be the lower
field ($e_L$ or $d_L$) rotation, as in this way the flavor violating $K$
or $U$ matrices appear explicitly in less experimentally constrained
$u$-quark or neutrino couplings (see also discussion in
Ref.~\cite{Aebischer:2015fzz}).  Our redefinitions are collected in
Table~\ref{tab:redef}.

Finally, Higgs boson interactions with fermions are affected by the
transition to the physical mass eigenstates both universally, due to
the change of Higgs-boson normalization in \eq{eq:Zh}, and in a flavor
dependent way, due to the modified relation in \eq{eq:modyuk} between
fermion masses and the Yukawa couplings. The Higgs-fermion-fermion
interaction Lagrangian in mass basis is,
\begin{eqnarray}
\mathcal{L}_{h\psi\psi} = &-& \bar{e} \, \left [ \frac{{M}_e}{v}
\left( 1-{1\over 4} C^{\varphi D}v^2+C^{\varphi\Box} v^2 \right) -
C^{e\varphi} \, \frac{v^2}{\sqrt{2}} \right  ]  P_R \, e \, h +
\mathrm{H.c.}  \nn
&-& \bar{u} \, \left [ \frac{{M}_u}{v}
\left( 1-{1\over 4} C^{\varphi D}v^2+C^{\varphi\Box} v^2 \right) -
C^{u\varphi} \, \frac{v^2}{\sqrt{2}} \right ]  P_R \, u \, h +
\mathrm{H.c.}  \nn
&-& \bar{d} \, \left [ \frac{{M}_d}{v}
\left( 1-{1\over 4} C^{\varphi D}v^2+C^{\varphi\Box} v^2 \right) -
C^{d\varphi} \, \frac{v^2}{\sqrt{2}} \right ] P_R \, d \, h +
\mathrm{H.c.} \;,
\end{eqnarray}
with the diagonal fermion mass matrices above, defined in \eq{eq:uu}.
Note that the dimension-5 operator in \eq{qnunu}, induces also a
Higgs-neutrino-neutrino vertex but this is highly suppressed since it
is proportional to neutrino masses.

\section{Gauge fixing and FP-ghosts in $R_\xi$-gauges}
\label{sec:gaugefix}

Compared to SM, the procedure of gauge fixing in SMEFT involves
additional features.  A consistent and convenient, for practical
purposes, choice of gauge fixing conditions and ghost sector should
fulfil the following requirements:
\begin{itemize}
\item Cancel the unwanted Goldstone-gauge boson bilinear mixing, as in
  SM.
\item Lead to SM-like propagators in terms of the effective mass basis
  parameters and fields.
\item Preserve the BRST invariance of the full Lagrangian in the
  presence of gauge fixing and ghost terms.
\end{itemize}

Let us notice that the gauge basis Lagrangian in terms of barred
couplings and fields, as obtained through \eq{eq:zgdef}, keeps the
same form up to rescaling factors.  For the dimension-4 terms it
reads,
\begin{align}
\mathcal{L}_{SM}^{(4)} & = -{1\over 4} Z_{g_s}^{-2}\bar G_{\mu\nu}^A
\bar G^{A\mu\nu}-{1\over 4} Z_{g}^{-2}\bar W_{\mu\nu}^I \bar
W^{I\mu\nu}-{1\over 4} Z_{g'}^{-2} \bar B_{\mu\nu} \bar B^{\mu\nu}
 \nn
& +(\bar D_\mu \varphi)^\dag (\bar D^\mu \varphi) + m^2 \varphi^\dag
 \varphi -{1\over 2}\lambda (\varphi^\dag \varphi)^2\nn
&+\ i(\bar l^\prime_L \slashed{\bar D} l^\prime_L + \bar e^\prime_R
 \slashed{\bar D} e^\prime_R + \bar q^\prime_L \slashed{\bar D}
 q^\prime_L + \bar u^\prime_R \slashed{\bar D} u^\prime_R +\bar
 d^\prime_R \slashed{\bar D} d^\prime_R )\nn
& - (\bar l^\prime_L \Gamma_e e^\prime_R \varphi + \bar q^\prime_L
 \Gamma_u u^\prime_R \tvp + \bar q^\prime_L \Gamma_d d^\prime_R
 \varphi + \mathrm{H.c.} )\;, \label{eq:smbar4}
\end{align}
while all higher dimensional operators remain unaffected at the
considered order.  Each term in the ``barred'' Lagrangian is still
manifestly $SU(3)\times SU(2)\times U(1)$ invariant, despite the
presence of $Z$-factors.  Therefore, we may equivalently use this
Lagrangian to gauge fix the theory.

Our choice for the gauge fixing term in the electroweak sector reads
\bea
\mathcal{L}_{GF} = - {1\over 2}\bm{F}^\top \bm{\hat{\xi}^{-1} F}\;,
\label{eq:lgfix0}
\eea
with the gauge fixing functionals $F^i$ defined through
\bea
\bm{F}= \left(\begin{array}{c} 
 F^1\\[2mm]
 F^2\\[2mm]
 F^3 \\[2mm]
 F^0 
\end{array}\right)
=
 \left(\begin{array}{c} 
   \partial_\mu  \bar W^{1\mu} \\[2mm]
    \partial_\mu \bar  W^{2\mu} \\[2mm]
   \partial_\mu  \bar W^{3\mu} \\ [2mm]
    \partial_\mu \bar B^\mu 
 \end{array}\right)
 -\frac{v\bm{\hat{\xi}}}{2}\left(\begin{array}{c}
   - i\bar g \frac{ \Phi^+ - \Phi^-}{\sqrt{2}}\\[2mm]
   \bar g \frac{\Phi^+ +  \Phi^-}{\sqrt{2}}\\[2mm]
    - \bar g Z_{G^0}^2  \Phi_0\\[2mm]
     \bar g' Z_{G^0}^2  \Phi_0
 \end{array}\right)
 \label{eq:lgfix}
\eea
and a $4\times 4$ \emph{symmetric} matrix ${\hat{\xi}}$ introduced as
\begin{align}
\hat{\bm{\xi}}= \left(\begin{array}{ccc} 
\xi_W& & 0 \\ 
&\xi_W& \\
0 & &\mathbb{X}\left(\begin{array}{cc} 
\xi_Z& \\ 
 &\xi_A \\ 
\end{array}\right)\mathbb{X}^\top\\
\end{array}\right)\,,
\end{align}
with $\mathbb{X}$ being the $2\times 2$ mixing matrix of the neutral
electroweak gauge bosons in \eq{eq:X}.

With such a choice in gauge basis, the transformations which
diagonalize and rescale the electroweak gauge and Goldstone bosons
also bring the gauge fixing term in a familiar form.  After
substituting the mass basis fields into \eq{eq:lgfix}, we arrive at
the expression
\bea
\mathcal{L}_{GF} = &-& {1\over \xi_W}(\partial^\mu W^+_\mu + i\xi_W
M_W G^+)(\partial^\nu W^-_\nu - i\xi_W M_W G^-)\nn
&-& {1\over 2\xi_Z}(\partial^\mu Z_\mu + \xi_Z M_Z G^0)^2 - {1\over
  2\xi_A}(\partial^\mu A_\mu)^2 \;,
\label{eq:lgfixm}
\eea
which looks identical to the SM one in the standard linear
$R_\xi$-gauges and has all terms required to eliminate the ``unwanted"
Goldstone-gauge mixing of \eq{eq:GEW}, through a total derivative. As
previously mentioned in Section~\ref{sec:goldstone}, such a standard
choice for $R_\xi$-gauges is possible since, in mass basis, all Wilson
coefficients of the ``unwanted" terms become absorbed in masses and
fields.

The gauge fixing conditions violate gauge invariance and we need to
introduce a ghost term in the Lagrangian to compensate and restore
(the more general) BRST invariance.  A convenient and consistent
choice for a ghost term takes the form
\bea
\mathcal{L}_{FP}  = \bm{\bar N}^\top \bm{\hat{\bm{E}}(\hat{\bm{M}}_F N)}\;,
\label{eq:lghost}
\eea
where the gauge basis ghost, anti-ghost fields are defined as
$N^i=(N^1,N^2,N^3,N^0)$, ${\bar N}^i=(\bar N^1,\bar N^2,\bar N^3,\bar
N^0)$, respectively and we have also introduced the \emph{symmetric}
$4\times 4$ matrix,
\bea
\bm{\hat{E}} &=& \left(\begin{array}{cc} 
\mathbb{1}_{2\times 2} & 0_{2\times 2}\\ 
0_{2\times 2} &  (\mathbb{X}^\top)^{-1} \mathbb{X}^{-1}\\
\end{array}\right).
\eea

The gauge fixing functionals $F^i$ chosen in \eq{eq:lgfix} are linear
in the fields and therefore the standard Faddeev-Popov (FP) treatment
with determinants applies\footnote{In the FP-treatment, it is clear
  that the matrix ${\hat{E}}$ factors out from the determinant as
  $\det(\bm{\hat{E}}\bm{\hat{M}_F})=\det(\bm{\hat{E}})\det(\bm{\hat{M}_F})$,
  affecting the path integral with an irrelevant constant
  factor.}. The explicit form of ${\hat M_F}$ can be always obtained
by performing an infinitesimal gauge transformation on $F^i$.
However, since we also wish to demonstrate the BRST invariance of the
SMEFT action we follow instead an equivalent derivation of $\hat{M}_F$
with the help of the BRST-operator, $\bm{s}$. It reads,
\bea
\hat M_F^{ij} N^j = \bm{s} {F^i} \;,
\label{eq:mfdef}
\eea
where \emph{lowercase} Latin indices run in the electroweak space
($\{i,j\}$=1,2,3,0).

Despite the presence of (constant) mixing matrices in the gauge fixing
functionals, the $\bm{s}$-operator transforms the fields included in
$F^i$, in a way identical to SM, as
\bea
\bm{s} \varphi &=& -i\bar g' Y \varphi~ N^0 -i\bar g {T^I} \varphi ~N^I \;, \nn
\bm{s} \varphi^\dag &=& +i\bar g' \varphi^\dag~Y N^0 +i\bar g
\varphi^\dag{T^I} ~N^I\;, \nn
\bm{s} \bar B_\mu &=&  \partial_\mu N^0\;, \nn
\bm{s} \bar W_\mu^I &=&  \partial_\mu N^I -\bar g \epsilon^{IJK}
\bar W_\mu^J N^K\;. \label{eq:slavnov}
\eea
Then,  $\hat M_F$ reads explicitly,
\bea
\bm{\hat M_F N}
&=& 
\partial^2\bm{N}
 + \bar g \overset{\leftarrow}{\partial^\mu} 
\left(\begin{array}{cccc}
 0 & - \bar W^3_\mu &  \bar W^2_\mu & 0 \\ 
  \bar W^3_\mu &  0&  -  \bar W^1_\mu & 0 \\ 
- \bar W^2_\mu &  \bar W^1_\mu &  0 & 0  \\ 
0 & 0 &0 & 0
\end{array}\right)\bm{N}\\
&+&\frac{v\bar g^2\hat{\bm{\xi}}}{4} \left(
\begin{array}{cccc}
H + v & \Phi^0 & \frac{ \Phi^+ + \Phi^- }{\sqrt{2}} & \frac{\bar g'}{\bar g}
\frac{ \Phi^+ + \Phi^- }{\sqrt{2}} \\[2mm]
- \Phi^0 & H + v & i\frac{ \Phi^+ - \Phi^- }{\sqrt{2}} & \frac{i\bar g'}{ \bar
  g} \frac{ \Phi^+ - \Phi^- }{\sqrt{2}} \\[2mm]
- Z_{G^0}^2 \frac{ \Phi^+ + \Phi^- }{\sqrt{2}} & - i Z_{G^0}^2 \frac{ \Phi^+ -
  \Phi^- }{\sqrt{2}} & Z_{G^0}^2(H + v ) & -\frac{\bar g'}{\bar g}
Z_{G^0}^2(H + v ) \\[2mm]
\frac{\bar g'}{ \bar g} Z_{G^0}^2 \frac{ \Phi^+ + \Phi^- }{\sqrt{2}} &
\frac{i\bar g'}{ \bar g} Z_{G^0}^2 \frac{ \Phi^+ - \Phi^-}{\sqrt{2}} &
-\frac{\bar g'}{ \bar g} Z_{G^0}^2 (H + v ) & \frac{\bar
  g^{'2}}{\bar g^2} Z_{G^0}^2(H + v )  \\ 
\end{array} 
\right) \bm{N}\nonumber
\eea

Once again, the chosen form of \eq{eq:lghost} with the presence of the
matrix $\hat E$, makes the transformation which diagonalizes the gauge
bosons kinetic terms and masses to diagonalize also ghost bilinear
terms. By defining ghost and anti-ghost fields in mass basis through
the relations
\begin{align}
 {1\over \sqrt{2}}( N^1 \mp i  N^2) = \eta^\pm ,&~~~~
 {1\over \sqrt{2}}( \bar N^1 \pm i \bar N^2) = \bar \eta^\pm    \;,
   \\ 
 \left(\begin{array}{c}
N^3 \\
N^0 \end{array}\right) =\mathbb{X}\left(\begin{array}{c}
 \eta^Z \\
 \eta^A \end{array}\right) ,&~~~~
\left(\begin{array}{c}
\bar N^3 \\
 \bar N^0 \end{array}\right)^\top =\left(\begin{array}{c}
 \bar \eta^Z \\
\bar \eta^A \end{array}\right)^\top \mathbb{X}^\top   \;,
\end{align}
all occurrences of the $\mathbb{X}$ matrix in bilinear ghost terms
become absorbed, leaving them in a canonical form with squared masses
$\xi_W M_W^2, ~\xi_Z M_Z^2$ and zero for the corresponding photon
ghost.  Again, the ghost propagators are SM-like (see
Appendix~\ref{app:propagators}). Nevertheless, corrections from higher
dimensional operators appear explicitly in ghost vertices as it was
also mentioned in \Ref{Hartmann:2015oia}.

The BRST invariance of the SMEFT action not including the gauge fixing
and ghost sector, follows immediately from its gauge invariance.
In order to establish BRST for the gauge fixing and ghost sector, as
well, we consider,
\bea
\bm{s}N^0 = 0~,~~\bm{s}N^I = {\bar g\over 2 } \epsilon^{IJK}N^J
N^K\;, \label{eq:N}\\ \bm{s}\bar N^i =
{F}^j(\bm{\hat{\xi}^{-1}\hat{E}^{-1}})^{ji}~.~~~~\label{eq:Nbar}
\eea
Using \eq{eq:mfdef} and \eq{eq:Nbar}, the property
$\bm{\hat{\xi}}^{-1}=(\bm{\hat{\xi}}^{-1})^\top$ and the relation
$\bm{s}{({\hat M}_F N)}=0$, which is associated with the nilpotency of
BRST, one obtains
\bea
\bm{s} \,\mathcal L_{GF} &=& -{1\over 2} \bm{s} \Big(F^i (\hat{\xi}^{-1})^{ij}
  F^j\Big)=-  F^i ({ \hat{\xi}^{-1}} )^{ij} (\bm{s}F^j) \nn
&=&-(\bm{s}\bar N^i) {\hat{E}}^{ij}{\hat{M}_F}^{jk}
N^k =- \bm{s}\Big({\bar N^i }{\hat{E}}^{ij} {\hat{M}_F}^{jk}
N^k\Big)= -\bm{s} \, \mathcal L_{FP}.
\eea
Hence, the full Lagrangian now remains invariant under BRST-symmetry
transformations.

As easily noticed, the BRST transformation on all gauge basis fields,
besides anti-ghosts, is identical to SM. Therefore, for this set of
fields it is nilpotent. The gauge fixing functionals $F^i$, although
modified by the presence of new (constant) mixing matrices, are still
linear functions of the same fields as in SM (\ie gauge and Goldstone
bosons). Thus, the BRST transformation for them is also nilpotent,
satisfying $\bm{s}^2 F^i = \bm{s}(M_F^{ij} N^j) = 0$, which can be
always verified explicitly. Finally, we note that the presence of
constant matrices in the transformation for anti-ghosts is in practice
irrelevant. This is because one can always introduce auxiliary fields
\cite{Lautrup:1967zz,Nakanishi:1966zz} in a suitable manner without
eventually affecting the gauge fixing and ghost terms. The choice
\bea
\mathcal{L}_{GF} = -\bm{B}^\top \bm{\hat{E} F}+ {1\over 2}\bm{B}^\top
\bm{\hat{E}\hat{\xi}\hat{E} B} \;,
\label{eq:lgfix0B}
\eea
is equivalent to \eq{eq:lgfix0} when the equations of motion are taken
for the auxiliary fields $B^i$. Changing only the transformation for
anti-ghosts, into $\bm{s}\bar N^i = B^i$ and introducing the new one
$\bm{s}B^i = 0$ for the auxiliary fields, one can verify that the
action remains BRST-invariant. Moreover, the BRST transformation on
all fields is now nilpotent, that is
\bea
\bm{s}^2=0.
\eea

In the QCD-sector, an analogous discussion of the $R_{\xi}$-gauges is
far more trivial. In terms of barred fields and couplings, the gauge
fixing and ghost terms read
\bea
\mathcal{L}_{GF}+\mathcal{L}_{FP} = -{1\over 2 \xi_G} F^A F^A +
\bar\eta_G^A M_F^{AB} \eta_G^B\,,
\eea
with simply, 
\bea
F^A &=& \partial_{\mu} g^{A\mu},\nn
M_F^{AB}\eta^B &=& \partial^2 \eta_G^A +
\bar{g}_s\overset{\leftarrow}{\partial_\mu} f^{ABC}g^{B\mu} \eta_G^C\,.
\eea

\section{Feynman rules and {\em Mathematica} implementation}
\label{sec:feyrul}

In Appendix~\ref{app:vert} we have collected the Feynman rules for
SMEFT propagators and interaction vertices in the $R_\xi$-gauges.
Most of the vertices are reasonably compact and for many processes
they can be readily used even for manual calculations.  We did not
display explicitly only the five and six gluon self-interactions as
they are, after symmetrizing in all Lorentz and color indices, very
long and it is unlikely that they can be used in any calculations
without the use of computer symbolic algebra programs.

Apart from the printed version, we have developed a publicly available
{\em Mathematica} code, {\tt SmeftFR}, calculating the same set of
Feynman rules with the use of {\tt FeynRules}
package~\cite{Alloul:2013bka}.  {\tt SmeftFR} is able to generate
automatically ``model files'' for {\tt FeynRules} for user-defined
subset of SMEFT operators (with numerical values of corresponding
Wilson coefficients defined in the WCxf
format~\cite{Aebischer:2017ugx}) and for chosen type of gauge fixing
conditions, further performing the field redefinitions described
earlier in the paper. The evaluated Feynman rules can be exported in
several commonly used formats (Latex, UFO~\cite{Degrande:2011ua},
FeynArts~\cite{Hahn:2000kx} and others), such that they can be
directly fed to other symbolic or numerical packages for high energy
physics calculations.  In addition, {\tt SmeftFR} provides set of
auxiliary programs performing extra manipulations atop {\tt
  FeynRules}--produced result, like optional correction of relative
sign between terms in 4-fermion vertices, correcting the expressions
for $B$ and/or $L$ lepton violating interactions and others. The
detailed package description and user manual can be found in
ref.~\cite{Dedes:2019uzs}.  The {\tt SmeftFR} package code can be
downloaded from {\tt \url{www.fuw.edu.pl/smeft}}.

Very recently it has appeared in the literature a {\em Mathematica}
program, called {\tt DsixTools}\cite{Celis:2017hod} that calculates
the Renormalization Group Equation (RGE) running of Wilson
coefficients for the operators listed in Tables~\ref{tab:no4ferm} and
\ref{tab:4ferm}. This code is complementary to our SMEFT code when
calculating renormalized amplitudes at leading and, up to
modifications, next to leading order in perturbation theory.

\section{Conclusions}
\label{sec:Con}

It is a central problem in particle physics today to categorize and
parametrize New Physics effects that are expected to arise by new
effective operators at some scale $\Lambda$.  In this article we
analyzed the structure of Standard Model Effective Field Theory
(SMEFT) including non-re\-nor\-ma\-li\-za\-ble operators up to
dimension 6.  For the first time in literature we derived the complete
set of Feynman rules for this theory quantized in linear
$R_\xi$-gauges.

More precisely, we started from the well known ``Warsaw" basis of
\Ref{Grzadkowski:2010es}, where the complete set of independent gauge
invariant $d\le 6$ operators is given, and identified the mass
eigenstate fields after Spontaneous Symmetry Breaking (SSB).
In achieving that goal, we performed constant and gauge invariant
field and coupling redefinitions in such a way that all physical and
unphysical fields possess canonical kinetic terms.
Furthermore, we constructed gauge fixing functionals which in mass
basis have a form of the linear $R_\xi$-gauges used routinely in the
SM loop-calculations.
A general set of different gauge fixing parameters for each gauge
field has been introduced, for completeness and for additional
cross-checks of the theory.

In order to restore the broken gauge symmetry after adding the gauge
fixing terms, a set of Faddeev-Popov ghosts has been introduced.
The ghost Lagrangian has been chosen such that the ghost propagators
again have the SM-like structure, while the effect of higher
dimensional operators appears explicitly only in their interaction
vertices.  We also proved that our SMEFT action preserves BRST
invariance and provide the reader with pertinent transformations in
section~\ref{sec:gaugefix}.

In summary, after establishing all steps described above, the bilinear
part of SMEFT Lagrangian and all, physical and unphysical, field
propagators expressed in terms of physical masses have exactly the
same structure as in the SM (although certain relations of masses and
couplings, such as the $\rho$-parameter for example, are modified by
the new operators). The effect of new $d=5$ and 6 operators {\em
  appears explicitly only in triple and higher multiplicity vertices,
  either as modifications of the SM ones or as genuine new
  interactions beyond the~SM}.

Within the mass basis considered here, we constructed the complete set
of Feynman rules in the linear $R_{\xi}$-gauges, not resorting to any
restriction such as CP- or baryon- lepton-number conservation.
The Feynman Rules for the total 380 vertices (not counting the
hermitian conjugate ones), which are about four times more than the SM
vertices, are given in Appendix~~\ref{app:vert}.
All Feynman rules were derived using the {\tt FeynRules} code and a
set of auxiliary programs created by the authors to perform field
redefinitions, various simplifications and an automatic translation to
Latex/axodraw format.  All propagators and vertices are listed
explicitly in the Appendix~\ref{app:vert} and can be recreated in
various symbolic computer formats with the use of a publicly available
{\em Mathematica} package, called \texttt{SmeftFR}, that can be
downloaded from\\[1mm]
\centerline{\tt \url{www.fuw.edu.pl/smeft}}\\[2mm]
The reader can consult {\tt SmeftFR} manual~\cite{Dedes:2019uzs} for
programming and installation details.

On the practical side, we believe that our SMEFT collection of Feynman
rules should significantly facilitate future phenomenological
analyses, saving time in deriving from scratch often lengthy
expressions in a complicated theory.
In addition, our Feynman rules help to avoid possible mistakes and
omissions of diagrams, which could easily happen when taking into
account only some parts of the full Lagrangian, as this is done in
many studies so far.  Furthermore, the publicly available {\tt
  SmeftFR} package that accompanies this article, can be used to
interface obtained Feynman rules to other high energy physics
computational computer programs, again streamlining the calculation of
future SMEFT physical predictions.

\section*{Acknowledgments}


AD would like to thank Apostolos Pilaftsis for illuminating
discussions about loop calculations in effective field theory.
The work of JR and WM is supported in part by the National Science
Centre, Poland, under research grants
DEC-2015/19/B/ST2/02848,
DEC-2015/18/M/ST2/00054
and DEC-2014/15/B/ST2/02157.  JR would also like to thank University
of Ioannina and CERN for hospitality during his visits.
AD and KS would like to thank University of Warsaw for hospitality.
KS acknowledges full financial support from Greek State Schoralships
Foundation (I.K.Y).  
\newpage

\appendix

\section{Feynman Rules}
\label{app:vert}
\subsection{General Notation}

In this Appendix we list the complete set of Feynman rules for SMEFT
in the physical (mass eigenstates) field basis and in
$R_\xi$-gauges. Our general conventions follow Peskin\&Schroeder
textbook\cite{Peskin:1995ev}.

In our notation for interaction vertices, indices and momentum for
each particle (external leg) carry a \textit{common} number label.
Indices of external particles appear explicitly in the diagrams but
momenta are suppressed for a better visual result. The convention for
number labels is displayed below for the four possible topologies of
SMEFT. Momenta are always considered to be incoming.\\[-10mm]
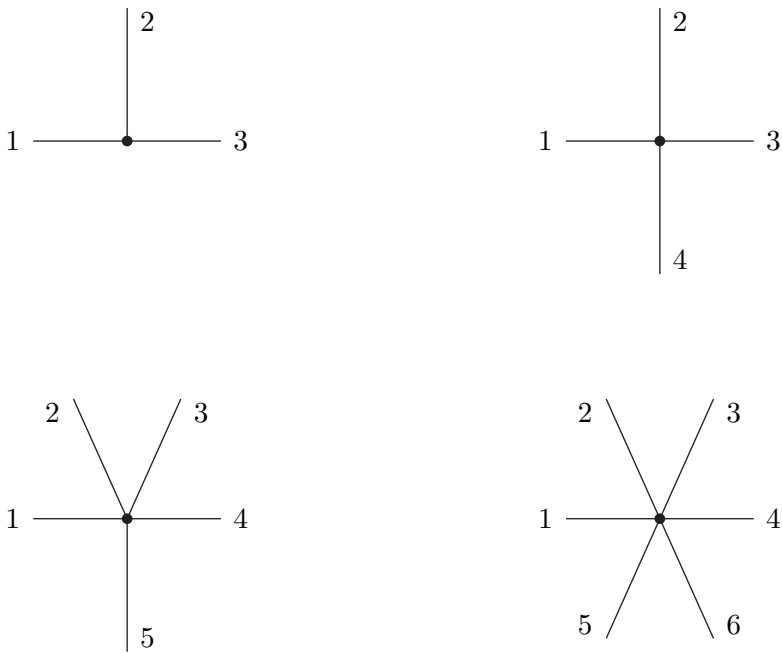
\begin{center}
\begin{tabular}{cp{3cm}c}
\begin{picture}(90,55)
\Line(5,0)(40,0)
\Text(0,0)[r]{$1$}
\Line(40,50)(40,0)
\Text(45,45)[l]{$2$}
\Line(75,0)(40,0)
\Text(80,0)[l]{$3$}
\Vertex(40,0){2}
\end{picture}
&&
\begin{picture}(90,110)
\Line(5,0)(40,0)
\Text(0,0)[r]{$1$}
\Line(40,50)(40,0)
\Text(45,45)[l]{$2$}
\Line(75,0)(40,0)
\Text(80,0)[l]{$3$}
\Line(40,-50)(40,0)
\Text(45,-45)[l]{$4$}
\Vertex(40,0){2}
\end{picture}
\\[10mm]
\begin{picture}(90,110)
\Line(5,0)(40,0)
\Text(0,0)[r]{$1$}
\Line(20,45)(40,0)
\Text(15,40)[r]{$2$}
\Line(60,45)(40,0)
\Text(65,40)[l]{$3$}
\Line(75,0)(40,0)
\Text(80,0)[l]{$4$}
\Line(40,-50)(40,0)
\Text(45,-45)[l]{$5$}
\Vertex(40,0){2}
\end{picture}
&&
\begin{picture}(90,110)
\Line(5,0)(40,0)
\Text(0,0)[r]{$1$}
\Line(20,45)(40,0)
\Text(15,40)[r]{$2$}
\Line(60,45)(40,0)
\Text(65,40)[l]{$3$}
\Line(75,0)(40,0)
\Text(80,0)[l]{$4$}
\Line(20,-45)(40,0)
\Text(15,-40)[r]{$5$}
\Line(60,-45)(40,0)
\Text(65,-40)[l]{$6$}
\Vertex(40,0){2}
\end{picture}
\end{tabular}  
\end{center}
\vskip 2cm

\noindent In addition to the notation defined in the main paper, we
use the following symbols:

\begin{center}
\begin{tabular}{lp{1cm}l}
\hline
Index type  & & Symbols \\
\hline
Flavor (generation) && $f_i,g_i$ \\
Spinor && $s_i$ \\
Color in triplet representation (quarks) && $m_i$ \\
Color in adjoint representation (gluons) && $a_i,b_i$\\
Lorentz && $\mu_i,\nu_i,\alpha_i,\beta_i,...$ \\
\hline
\end{tabular}\\[2mm]
\end{center}

\noindent 
and $\eta_{\mu\nu}$ denotes the Minkowski metric tensor with
signature $(+,-,-,-)$.

The Feynman rules listed in this Appendix were generated fully
automatically by a specialized {\em Mathematica} code directly
producing Latex output. In order to avoid possible misprints, in
manual edits of this output we kept our changes minimal. It was also
difficult to implement in such a {\em Mathematica} to Latex translator
the proper positioning of Lorentz indices. In the expressions of this
Appendix one should assume that repeating Lorentz indices are always
contracted in a covariant way, even if they are not
subscript-superscript pairs.

\subsection{Conventions for fermions}
\label{app:fermion}

Contrary to the bosonic sector, amplitude calculations in the
fermionic sector of SMEFT require additional care. The effects which
complicate the picture are discussed
elsewhere\cite{Paraskevas:2018}. For practical use, the following
instructions can be useful:
\begin{itemize}
\item Our conventions are consistent with Peskin\&Schroeder
  textbook. One can always use the full toolbox offered there,
  including spin sums and other identities. Our phase convention for
  charge conjugation is adapted to
  Bjorken\&Drell~\cite{Bjorken:1965zz}, as followed
  by~\cite{Grzadkowski:2010es}.
\item {Neutrino fields are in general massive in SMEFT, and as such
  have to be represented by Majorana fermions.  Amplitude calculations
  involving Majorana neutrinos require a generalization of the
  standard formalism developed by Denner \textit{et
    al.}~\cite{Denner:1992vza, Denner:1992me}. }
\item When four-fermion {vertices} are relevant an even larger
  generalization is required. The extension of the standard formalism
  dealing properly with amplitudes involving four-fermion vertices
    has been proposed recently in Ref.~\cite{Paraskevas:2018}.
\item {For the generalized formalisms it is useful to introduce the
  notion of the \textit{fermion-flow}\cite{Denner:1992vza}}.  In our
  propagators and vertices, arrows on fermion lines always denote such
  fermion flow. For the usual Dirac fields, fermion-number and
  fermion-flow arrows have the same direction and there is no
  distinction between the two. For charge-conjugate fields (appearing
  in $B$,$L$-violating processes), they have the opposite direction.
  For Majorana neutrinos, fermion number is meaningless and one uses
  fermion-flow instead. We completely suppress fermion-number arrows
  in all cases since, if needed, they can be trivially reproduced with
  these remarks.
\end{itemize}

\subsubsection{Conventions for fermion-number violation and neutrinos.}

In the chiral representation, the charge conjugation unitary
matrix reads,
\begin{align} 
&\cc =-i \gamma^2 \gamma^0 ~,\\
\cc^\dag = &\cc^{-1}= \cc^{\trp} = -\cc ~.
\end{align}
The $\cc$-matrix can associate the Dirac four-component (commuting)
spinors, as
\begin{align}
u(p,\bar s)=\cc\bar v^\trp (p,\bar s)~~,~~~ v(p,\bar s)=\cc\bar u^\trp
(p,\bar s)~. \label{ccspinors}
\end{align}
It can be further used to define the charge-conjugate fields as,
\begin{align}
\psi^c = \cc\bar \psi^\trp ~~,~~~ \bar\psi^c=\psi^\trp
\mathbb{C}~.  \label{psicc}
\end{align} 
Obviously, these fields do not represent new degrees of freedom in the
spectrum. Nevertheless, introducing them within the formalism of
\cite{Denner:1992vza} simplifies calculations of fermion-number
violating effects, in a remarkable way.

In our Feynman rules, neutrinos are considered to be Majorana
particles.  This choice can describe consistently all processes of
SMEFT.  We adopt the phase convention,
\bea
\nu = \nu^c = \cc \bar\nu^\trp ~.
\eea
All Feynman rules involving neutrinos listed in this Appendix have
been properly symmetrized in their indices, taking also into account
the Majorana condition when relevant. In addition, they require using
the real-scalar type combinatorics for the diagram multiplicities (for
example, an additional $1/2$ factor for a Majorana neutrino loop,
besides the usual minus sign).

\begin{table}[tb]
\centering
\begin{tabular}{|c|c c c|}
\hline
\rule[-1ex]{0pt}{2.5ex} &Massive Majorana &$\longrightarrow$ &
Massless Weyl \\ [0.5ex]
\hline
\rule[0.5ex]{0pt}{2.5ex} Mass &$m_{\nu}$ & $\longrightarrow$ & 0 \\[0.5ex] 
\rule[0.5ex]{0pt}{2.5ex} PMNS &$ U $
&$\longrightarrow $ & $\mathbb{1}$ \\[0.5ex]
\rule[0.5ex]{0pt}{2.5ex} Propagator &${i \over \slashed{p}-m}$
&$\longrightarrow$ & ${i \over \slashed{p}}P_R$ \\[0.5ex]
\rule[0.5ex]{0pt}{2.5ex} Vertices &$\delta_{f_i
  f_j}(\gamma^\mu\gamma^5)_{s_is_j}$ &$\longrightarrow$& $-\delta_{f_i
  f_j}(\gamma^\mu P_L)_{s_is_j}$ \\[0.5ex]
\rule[0.5ex]{0pt}{2.5ex} Vertices & $(\gamma^\mu P_R)_{s_is_j},
(\slashed{p}P_R)_{s_is_j}$ &$\longrightarrow$& 0 \\[0.5ex]
\rule[0.5ex]{0pt}{2.5ex} Combinatorics & Real-scalar
&$\longrightarrow$ & Charged-scalar \\[0.5ex] \hline
\end{tabular}
\caption{\sf Conversion of neutrino Feynman rules from the massive
  Majorana case to the massless Weyl case. The pair of index-labels
  $i,j$ above refers to neutrino legs in vertices. In the
  4$\nu$-vertex one should further eliminate the $(\dots)_{s_1s_3}$
  terms for Weyl spinors.}
\label{tab:conv}
\end{table}

For many applications where effects of the neutrino masses are
negligible it may be easier to work in the neutrino massless-limit
with Weyl spinors.  Weyl Feynman rules can be derived from the
Majorana rules given in the following sections of our Appendix using
the transformations of Table~\ref{tab:conv}.

\subsubsection{Ordering conventions for four-fermion vertices.}

The vertices have been extracted from the interaction Hamiltonian
($\mathcal{H}_{int}=-\mathcal{L}_{int}$) as,
\begin{align}
~ \bra{0} T[-i\mathcal{H}_{int}] b_4^\dag d_3^\dag b_2^\dag
  d_1^\dag\ket{0} & = i\hat{\mathbb{\Gamma}}_{1'2'3'4'}~ \bra{0} (
  \bar \psi_{1'} \psi_{2'} \bar \psi_{3'} \psi_{4'}) b_4^\dag d_3^\dag
  b_2^\dag d_1^\dag\ket{0}\Big|{\begin{array}{c}   \\
      \textnormal{\scriptsize{All possible}} \\[-2mm]
      \textnormal{\scriptsize{contractions}}
\end{array} } \nonumber \\
&= i\mathbb{\Gamma}_{1234} \bar v_1 u_2 \bar v_3 u_4\;,
\end{align}
where $b^\dagger (d^\dagger)$ are creation operators for particles
(antiparticles) with obvious modifications for neutrino vertices
  (\ie $d^\dag =b^\dag$) and $B$,$L$-violating {interactions} ($d^\dag
  \to b^\dag$ for $\bar\psi\to \bar\psi^c$), recalling that momenta
are always considered incoming. Above, $\hat{\mathbb{\Gamma}}$ is the
initial (unsymmetrized) Lagrangian coupling in mass basis and
$i\mathbb{\Gamma}$ is the analytic expression for the (fully
symmetrized) vertex, both displayed here in a general tensor form. The
tensor numbers represent indices from the set,
$$i=\{f_i,s_i,m_i\},$$
which are relevant for each case.  For example, the $4e$-vertex of
SMEFT reads explicitly,
$$i\mathbb{\Gamma}_{1234} \equiv 2i (C^{ll}_{f_1f_2f_3f_4} (\gamma_\mu
P_L)_{s_1s_2} (\gamma^\mu P_L)_{s_3s_4}-C^{ll}_{f_1f_4f_3f_2}
(\gamma_\mu P_L)_{s_1s_4} (\gamma^\mu P_L)_{s_3s_2})+\dots \;,$$
where ellipses denote contribution from other coefficients.

\subsection{Propagators in the $R_\xi$-gauges}
\label{app:propagators}

\begin{tabular}{lp{10mm}l}
\begin{picture}(80,30)
\Photon(0,10)(80,10){3}{4} \Line(40,10)(33,17) \Line(40,10)(33,3)
\Text(0,0)[r]{$\mu$} \Text(80,0)[l]{$\nu$} \Text(40,25)[c]{$W^\pm$}
\end{picture}
&&
\raisebox{10\unitlength}{
\begin{minipage}{0.75\linewidth}
\begin{eqnarray*}
-{i\over k^2- M_W^2} \Big[\eta^{\mu\nu}-(1-\xi_W){k^\mu k^\nu \over
    k^2 -\xi_W M_W^2}\Big]
\end{eqnarray*}
\end{minipage}
}
\\[3mm]
\begin{picture}(80,30)
\Photon(0,10)(80,10){3}{4}
\Text(0,0)[r]{$\mu$}
\Text(80,0)[l]{$\nu$}
\Text(40,25)[c]{$Z^0$}
\end{picture}
&&
\raisebox{10\unitlength}{
\begin{minipage}{0.75\linewidth}
\begin{eqnarray*}
 -{i\over k^2- M_Z^2} \Big[\eta^{\mu\nu}-(1-\xi_Z){k^\mu k^\nu \over
     k^2 -\xi_Z M_Z^2}\Big]
\end{eqnarray*}
\end{minipage}
} \\[3mm]
\begin{picture}(80,30)
\Photon(0,10)(80,10){3}{4}
\Text(0,0)[r]{$\mu$}
\Text(80,0)[l]{$\nu$}
\Text(40,25)[c]{$A^0$}
\end{picture}
&&
\raisebox{10\unitlength}{
\begin{minipage}{0.75\linewidth}
\begin{eqnarray*}
-{i\over k^2} \Big[\eta^{\mu\nu}-(1-\xi_A){k^\mu k^\nu \over k^2
  }\Big]
\end{eqnarray*}
\end{minipage}
}
\\[3mm]
\begin{picture}(80,30)
\Photon(0,10)(80,10){3}{4}
\Text(0,0)[r]{$\mu,a$}
\Text(80,0)[l]{$\nu,b$}
\Text(40,25)[c]{$g$}
\end{picture}
&&
\raisebox{10\unitlength}{
\begin{minipage}{0.75\linewidth}
\begin{eqnarray*}
-{i\delta_{ab}\over k^2} \Big[\eta^{\mu\nu}-(1-\xi_G){k^\mu k^\nu
    \over k^2 }\Big]
\end{eqnarray*}
\end{minipage}
}
\\[3mm]
\begin{picture}(80,30)
\ZigZag(0,10)(80,10){3}{4}
\Text(0,0)[r]{$$}
\Text(80,0)[l]{$$}
\Text(40,25)[c]{$\eta^\pm$}
\end{picture}
&&
\raisebox{10\unitlength}{
\begin{minipage}{0.75\linewidth}
  \begin{eqnarray*}
  ~~~~~~~~~~~~~~~~~~~~~  -{i\over k^2-\xi_W M_W^2} ~~~~~~~~~\longrightarrow \wick{\c{\eta^{\pm}}\c{\bar\eta^{\pm}}}
  \end{eqnarray*}
\end{minipage}
}
\\
\end{tabular}

\noindent
Note that $\psi$ above stands for the Dirac fields of the theory,
$\psi=e,u,d$. Color {indices}, $m_i$, are relevant only for
quarks. Charge-conjugate fields $\psi^c$ are related to
$B$,$L$-violating vertices and do not represent new fields in SMEFT.
Momentum and ghost-number flow from left to right.


\bigskip
\bigskip
\bigskip
\subsection{Lepton--gauge vertices}

\noindent \begin{tabular}{cp{1mm}l}
\begin{picture}(90,55)(0,15)
\ArrowLine(40,0)(5,0)
\Text(0,0)[r]{$\nu^{f_1}$}
\ArrowLine(40,50)(40,0)
\Text(45,45)[l]{$\nu^{f_2}$}
\Photon(75,0)(40,0){3}{4}
\Text(80,0)[l]{$Z^0_{\mu_3}$}
\Vertex(40,0){2}
\end{picture}
&&
\begin{minipage}[c]{0.8\linewidth}
\begin{dmath*}
 \nolinebreak{+ \frac12 i \sqrt{{{\bar g}}^2 + {{\bar g'}{}}^2} \delta_{f_1 f_2} \gamma^{\mu_3} \gamma^5 + \frac{i {{\bar g}} {{\bar g'}{}} v^2}{2 \sqrt{{{\bar g}}^2 + {{\bar g'}{}}^2}} \delta_{f_1 f_2} C^{{ \varphi WB}} \gamma^{\mu_3} \gamma^5 }
 \linebreak
 \nolinebreak{+ \frac12 i v^2 \sqrt{{{\bar g}}^2 + {{\bar g'}{}}^2} \Big[U_{{g_2} f_2} U_{{g_1} f_1}^* C^{{ \varphi l1}}_{{g_1} {g_2}} \gamma^{\mu_3} P_L - U_{{g_1} f_1} U_{{g_2} f_2}^* C^{{ \varphi l1}}_{{g_2} {g_1}} \gamma^{\mu_3} P_R
 \Big]}
 \linebreak
 \nolinebreak{ - \frac12 i v^2 \sqrt{{{\bar g}}^2 + {{\bar g'}{}}^2}\Big[ U_{g_2 f_2} U_{g_1 f_1}^* C^{ \varphi l3}_{g_1 g_2} \gamma^{\mu_3} P_L- U_{{g_1} f_1} U_{g_2 f_2}^* C^{{ \varphi l3}}_{g_2 g_1} \gamma^{\mu_3} P_R\Big]}
\end{dmath*}

\end{minipage}
\end{tabular}

\bigskip

\noindent \begin{tabular}{cp{1mm}l}
\begin{picture}(90,55)(0,15)
\ArrowLine(40,0)(5,0)
\Text(0,0)[r]{$ e^{f_1}$}
\ArrowLine(40,50)(40,0)
\Text(45,45)[l]{$e^{f_2}$}
\Photon(75,0)(40,0){3}{4}
\Text(80,0)[l]{$A^0_{\mu_3}$}
\Vertex(40,0){2}
\end{picture}
&&
\begin{minipage}[c]{0.8\linewidth}
\begin{dmath*}
 + \frac{i {{\bar g}} {{\bar g'}{}}}{\sqrt{{{\bar g}}^2 + {{\bar g'}{}}^2}} \delta_{f_1 f_2} \gamma^{\mu_3} - \frac{i {{\bar g}}^2 {{\bar g'}{}}^2 v^2}{\left({{\bar g}}^2 + {{\bar g'}{}}^2\right)^{3/2}} \delta_{f_1 f_2} C^{{ \varphi WB}} \gamma^{\mu_3} + \frac{\sqrt2 {{\bar g'}{}} v}{\sqrt{{{\bar g}}^2 + {{\bar g'}{}}^2}} p_3^{\nu } \left(C^{{eW}*}_{f_2 f_1} \sigma^{\mu_3 \nu } P_L  + C^{{eW}}_{f_1 f_2} \sigma^{\mu_3 \nu } P_R \right) - \frac{\sqrt2 {{\bar g}} v}{\sqrt{{{\bar g}}^2 + {{\bar g'}{}}^2}} p_3^{\nu } \left(C^{{eB}*}_{f_2 f_1} \sigma^{\mu_3 \nu } P_L  + C^{{eB}}_{f_1 f_2} \sigma^{\mu_3 \nu } P_R \right)
\end{dmath*}

\end{minipage}
\end{tabular}

\bigskip

\noindent \begin{tabular}{cp{1mm}l}
\begin{picture}(90,55)(0,15)
\ArrowLine(40,0)(5,0)
\Text(0,0)[r]{$ e^{f_1}$}
\ArrowLine(40,50)(40,0)
\Text(45,45)[l]{$e^{f_2}$}
\Photon(75,0)(40,0){3}{4}
\Text(80,0)[l]{$Z^0_{\mu_3}$}
\Vertex(40,0){2}
\end{picture}
&&
\begin{minipage}[c]{0.8\linewidth}
\begin{dmath*}
 - \frac{i}{2 \sqrt{{{\bar g}}^2 + {{\bar g'}{}}^2}} \delta_{f_1 f_2} \left(\left({{\bar g'}{}}^2 - {{\bar g}}^2\right) \gamma^{\mu_3} P_L + 2 {{\bar g'}{}}^2 \gamma^{\mu_3} P_R\right) + \frac{i {{\bar g}} {{\bar g'}{}} v^2}{2 \left({{\bar g}}^2 + {{\bar g'}{}}^2\right)^{3/2}} \delta_{f_1 f_2} C^{{ \varphi WB}} \left(\left({{\bar g'}{}}^2 - {{\bar g}}^2\right) \gamma^{\mu_3} P_L - 2 {{\bar g}}^2 \gamma^{\mu_3} P_R\right) + \frac{\sqrt2 {{\bar g}} v}{\sqrt{{{\bar g}}^2 + {{\bar g'}{}}^2}} p_3^{\nu } \left(C^{{eW}*}_{f_2 f_1} \sigma^{\mu_3 \nu } P_L  + C^{{eW}}_{f_1 f_2} \sigma^{\mu_3 \nu } P_R \right) + \frac{\sqrt2 {{\bar g'}{}} v}{\sqrt{{{\bar g}}^2 + {{\bar g'}{}}^2}} p_3^{\nu } \left(C^{{eB}*}_{f_2 f_1} \sigma^{\mu_3 \nu } P_L  + C^{{eB}}_{f_1 f_2} \sigma^{\mu_3 \nu } P_R \right) + \frac12 i v^2 \sqrt{{{\bar g}}^2 + {{\bar g'}{}}^2} C^{{ \varphi l1}}_{f_1 f_2} \gamma^{\mu_3} P_L + \frac12 i v^2 \sqrt{{{\bar g}}^2 + {{\bar g'}{}}^2} C^{{ \varphi l3}}_{f_1 f_2} \gamma^{\mu_3} P_L + \frac12 i v^2 \sqrt{{{\bar g}}^2 + {{\bar g'}{}}^2} C^{{ \varphi e}}_{f_1 f_2} \gamma^{\mu_3} P_R
\end{dmath*}

\end{minipage}
\end{tabular}

\bigskip

\noindent \begin{tabular}{cp{1mm}l}
\begin{picture}(90,55)(0,15)
\ArrowLine(40,0)(5,0)
\Text(0,0)[r]{$\nu^{f_1}$}
\ArrowLine(40,50)(40,0)
\Text(45,45)[l]{$e^{f_2}$}
\Photon(75,0)(40,0){3}{4}
\Line(57.,0)(63.,-6.)
\Line(57.,0)(63.,6.)
\Text(80,0)[l]{$W^+_{\mu_3}$}
\Vertex(40,0){2}
\end{picture}
&&
\begin{minipage}[c]{0.8\linewidth}
\begin{dmath*}
 - \frac{i {{\bar g}}}{\sqrt2} U_{f_2 f_1}^* \gamma^{\mu_3} P_L - 2 v p_3^{\nu } U_{{g_1} f_1}^* C^{{eW}}_{{g_1} f_2} \sigma^{\mu_3 \nu } P_R  - \frac{i {{\bar g}} v^2}{\sqrt2} U_{{g_1} f_1}^* C^{{ \varphi l3}}_{{g_1} f_2} \gamma^{\mu_3} P_L
\end{dmath*}

\end{minipage}
\end{tabular}

\bigskip

\noindent \begin{tabular}{cp{1mm}l}
\begin{picture}(90,100)(0,50)
\ArrowLine(40,50)(5,50)
\Text(0,50)[r]{$ e^{f_1}$}
\ArrowLine(40,100)(40,50)
\Text(45,95)[l]{$e^{f_2}$}
\Photon(75,50)(40,50){3}{4}
\Line(57.,50.)(63.,44.)
\Line(57.,50.)(63.,56.)
\Text(80,50)[l]{$W^+_{\mu_3}$}
\Photon(40,0)(40,50){3}{4}
\Line(40.,25.)(34.,19.)
\Line(40.,25.)(46.,19.)
\Text(45,5)[l]{$W^-_{\mu_4}$}
\Vertex(40,50){2}
\end{picture}
&&
\begin{minipage}[c]{0.8\linewidth}
\begin{dmath*}
 + \sqrt2 {{\bar g}} v \left(\sigma^{\mu_3 \mu_4} P_L  C^{{eW}*}_{f_2 f_1} + C^{{eW}}_{f_1 f_2} \sigma^{\mu_3 \mu_4} P_R \right)
\end{dmath*}

\end{minipage}
\end{tabular}

\bigskip

\noindent \begin{tabular}{cp{1mm}l}
\begin{picture}(90,100)(0,50)
\ArrowLine(40,50)(5,50)
\Text(0,50)[r]{$ e^{f_1}$}
\ArrowLine(40,100)(40,50)
\Text(45,95)[l]{$\nu^{f_2}$}
\Photon(75,50)(40,50){3}{4}
\Text(80,50)[l]{$A^0_{\mu_3}$}
\Photon(40,0)(40,50){3}{4}
\Line(40.,25.)(34.,19.)
\Line(40.,25.)(46.,19.)
\Text(45,5)[l]{$W^-_{\mu_4}$}
\Vertex(40,50){2}
\end{picture}
&&
\begin{minipage}[c]{0.8\linewidth}
\begin{dmath*}
 + \frac{2 {{\bar g}} {{\bar g'}{}} v}{\sqrt{{{\bar g}}^2 + {{\bar g'}{}}^2}} U_{{g_1} f_2} \sigma^{\mu_3 \mu_4} P_L  C^{{eW}*}_{{g_1} f_1}
\end{dmath*}

\end{minipage}
\end{tabular}

\bigskip

\noindent \begin{tabular}{cp{1mm}l}
\begin{picture}(90,100)(0,50)
\ArrowLine(40,50)(5,50)
\Text(0,50)[r]{$ e^{f_1}$}
\ArrowLine(40,100)(40,50)
\Text(45,95)[l]{$\nu^{f_2}$}
\Photon(75,50)(40,50){3}{4}
\Line(57.,50.)(63.,44.)
\Line(57.,50.)(63.,56.)
\Text(80,50)[l]{$W^-_{\mu_3}$}
\Photon(40,0)(40,50){3}{4}
\Text(45,5)[l]{$Z^0_{\mu_4}$}
\Vertex(40,50){2}
\end{picture}
&&
\begin{minipage}[c]{0.8\linewidth}
\begin{dmath*}
 - \frac{2 {{\bar g}}^2 v}{\sqrt{{{\bar g}}^2 + {{\bar g'}{}}^2}} U_{{g_1} f_2} \sigma^{\mu_3 \mu_4} P_L  C^{{eW}*}_{{g_1} f_1}
\end{dmath*}

\end{minipage}
\end{tabular}

\bigskip

\bigskip
\bigskip
\bigskip

\subsection{Lepton--Higgs--gauge vertices}

\noindent \begin{tabular}{cp{1mm}l}
\begin{picture}(90,55)(0,15)
\ArrowLine(40,0)(5,0)
\Text(0,0)[r]{$ e^{f_1}$}
\ArrowLine(40,50)(40,0)
\Text(45,45)[l]{$e^{f_2}$}
\DashLine(75,0)(40,0){3}
\Text(80,0)[l]{$G^0$}
\Vertex(40,0){2}
\end{picture}
&&
\begin{minipage}[c]{0.8\linewidth}
\begin{dmath*}
 + \frac1{v} \delta_{f_1 f_2} m_{l_{f_1}} \gamma^5 - \frac{v}4 \delta_{f_1 f_2} C^{{ \varphi D}} m_{l_{f_1}} \gamma^5 - v \slashed{p}_3 P_L C^{{ \varphi l1}}_{f_1 f_2} - v \slashed{p}_3 P_L C^{{ \varphi l3}}_{f_1 f_2} - v \slashed{p}_3 P_R C^{{ \varphi e}}_{f_1 f_2}
\end{dmath*}

\end{minipage}
\end{tabular}

\bigskip

\noindent \begin{tabular}{cp{1mm}l}
\begin{picture}(90,55)(0,15)
\ArrowLine(40,0)(5,0)
\Text(0,0)[r]{$ e^{f_1}$}
\ArrowLine(40,50)(40,0)
\Text(45,45)[l]{$e^{f_2}$}
\DashLine(75,0)(40,0){3}
\Text(80,0)[l]{$h$}
\Vertex(40,0){2}
\end{picture}
&&
\begin{minipage}[c]{0.8\linewidth}
\begin{dmath*}
 - \frac{i}{v} \delta_{f_1 f_2}  m_{l_{f_1}} - i v \delta_{f_1 f_2}  C^{{ \varphi \Box}} m_{l_{f_1}} + \frac{i v}4 \delta_{f_1 f_2}  C^{{ \varphi D}} m_{l_{f_1}} + \frac{i v^2}{\sqrt2} \left(P_L C^{{e \varphi }*}_{f_2 f_1} + P_R C^{{e \varphi }}_{f_1 f_2}\right)
\end{dmath*}

\end{minipage}
\end{tabular}

\bigskip

\noindent \begin{tabular}{cp{1mm}l}
\begin{picture}(90,55)(0,15)
\ArrowLine(40,0)(5,0)
\Text(0,0)[r]{$\nu^{f_1}$}
\ArrowLine(40,50)(40,0)
\Text(45,45)[l]{$e^{f_2}$}
\DashArrowLine(75,0)(40,0){3}
\Text(80,0)[l]{$G^+$}
\Vertex(40,0){2}
\end{picture}
&&
\begin{minipage}[c]{0.8\linewidth}
\begin{dmath*}
 + \frac{i \sqrt2}{v} U_{f_2 f_1}^* \left(m_{\nu_{f_1}} P_L - m_{l_{f_2}} P_R\right) + i \sqrt2 v \slashed{p}_3 P_L U_{{g_1} f_1}^* C^{{ \varphi l3}}_{{g_1} f_2}
\end{dmath*}

\end{minipage}
\end{tabular}

\bigskip

\noindent \begin{tabular}{cp{1mm}l}
\begin{picture}(90,55)(0,15)
\ArrowLine(40,0)(5,0)
\Text(0,0)[r]{$\nu^{f_1}$}
\ArrowLine(40,50)(40,0)
\Text(45,45)[l]{$\nu^{f_2}$}
\DashLine(75,0)(40,0){3}
\Text(80,0)[l]{$G^0$}
\Vertex(40,0){2}
\end{picture}
&&
\begin{minipage}[c]{0.8\linewidth}
\begin{dmath*}
-{2 \over v} \delta_{f_1 f_2} m_{\nu_{f_1}}\gamma^5
 \nolinebreak{- v \Big[U_{{g_2} f_2} U_{{g_1} f_1}^* C^{{ \varphi l1}}_{{g_1} {g_2}}\slashed{p}_3 P_L - U_{{g_1} f_1} U_{{g_2} f_2}^* C^{{ \varphi l1}}_{{g_2} {g_1}}\slashed{p}_3 P_R \Big] }
 \linebreak
 \nolinebreak{+ v \Big[U_{{g_2} f_2}  U_{{g_1} f_1}^* C^{{ \varphi l3}}_{{g_1} {g_2}} \slashed{p}_3 P_L - U_{{g_1} f_1}  U_{{g_2} f_2}^* C^{{ \varphi l3}}_{{g_2} {g_1}} \slashed{p}_3 P_R\Big]}
\end{dmath*}

\end{minipage}
\end{tabular}

\bigskip

\noindent \begin{tabular}{cp{1mm}l}
\begin{picture}(90,55)(0,15)
\ArrowLine(40,0)(5,0)
\Text(0,0)[r]{$ \nu^{f_1}$}
\ArrowLine(40,50)(40,0)
\Text(45,45)[l]{$\nu^{f_2}$}
\DashLine(75,0)(40,0){3}
\Text(80,0)[l]{$h$}
\Vertex(40,0){2}
\end{picture}
&&
\begin{minipage}[c]{0.8\linewidth}
\begin{dmath*}
 - \frac{2i}{v} \delta_{f_1 f_2}  m_{\nu_{f_1}} 
\end{dmath*}

\end{minipage}
\end{tabular}

\bigskip

\noindent \begin{tabular}{cp{1mm}l}
\begin{picture}(90,100)(0,50)
\ArrowLine(40,50)(5,50)
\Text(0,50)[r]{$ (e^c)^{f_1}$}
\ArrowLine(40,100)(40,50)
\Text(45,95)[l]{$ e^{f_2}$}
\DashArrowLine(75,50)(40,50){3}
\Text(80,50)[l]{$G^+$}
\DashArrowLine(40,0)(40,50){3}
\Text(45,5)[l]{$G^+$}
\Vertex(40,50){2}
\end{picture}
&&
\begin{minipage}[c]{0.8\linewidth}
\begin{dmath*}
-  {4i\over v^2} ~m_{\nu_{g_1}}  U^*_{f_1 g_1} U^*_{f_2 g_1}  P_L
\end{dmath*}

\end{minipage}
\end{tabular}

\bigskip

\noindent \begin{tabular}{cp{1mm}l}
\begin{picture}(90,100)(0,50)
\ArrowLine(40,50)(5,50)
\Text(0,50)[r]{$ e^{f_1}$}
\ArrowLine(40,100)(40,50)
\Text(45,95)[l]{$e^{f_2}$}
\DashArrowLine(75,50)(40,50){3}
\Text(80,50)[l]{$G^+$}
\DashArrowLine(40,0)(40,50){3}
\Text(45,5)[l]{$G^-$}
\Vertex(40,50){2}
\end{picture}
&&
\begin{minipage}[c]{0.8\linewidth}
\begin{dmath*}
-  {4i\over v^2} ~m_{\nu_{g_1}}  U^*_{f_1 g_1} U^*_{f_2 g_1}  P_L
\end{dmath*}

\end{minipage}
\end{tabular}

\bigskip

\noindent \begin{tabular}{cp{1mm}l}
\begin{picture}(90,100)(0,50)
\ArrowLine(40,50)(5,50)
\Text(0,50)[r]{$ e^{f_1}$}
\ArrowLine(40,100)(40,50)
\Text(45,95)[l]{$e^{f_2}$}
\DashLine(75,50)(40,50){3}
\Text(80,50)[l]{$G^0$}
\DashLine(40,0)(40,50){3}
\Text(45,5)[l]{$G^0$}
\Vertex(40,50){2}
\end{picture}
&&
\begin{minipage}[c]{0.8\linewidth}
\begin{dmath*}
 + \frac{i v}{\sqrt2} \left(P_L C^{{e \varphi }*}_{f_2 f_1} + P_R C^{{e \varphi }}_{f_1 f_2}\right)
\end{dmath*}

\end{minipage}
\end{tabular}

\bigskip

\noindent \begin{tabular}{cp{1mm}l}
\begin{picture}(90,100)(0,50)
\ArrowLine(40,50)(5,50)
\Text(0,50)[r]{$ e^{f_1}$}
\ArrowLine(40,100)(40,50)
\Text(45,95)[l]{$e^{f_2}$}
\DashLine(75,50)(40,50){3}
\Text(80,50)[l]{$G^0$}
\DashLine(40,0)(40,50){3}
\Text(45,5)[l]{$h$}
\Vertex(40,50){2}
\end{picture}
&&
\begin{minipage}[c]{0.8\linewidth}
\begin{dmath*}
 + \frac{v}{\sqrt2} \left(P_L C^{{e \varphi }*}_{f_2 f_1} - P_R C^{{e \varphi }}_{f_1 f_2}\right) -  \left(\slashed{p}_3 P_L - \slashed{p}_4 P_L\right) C^{{ \varphi l1}}_{f_1 f_2} -  \left(\slashed{p}_3 P_L - \slashed{p}_4 P_L\right) C^{{ \varphi l3}}_{f_1 f_2} -  \left(\slashed{p}_3 P_R - \slashed{p}_4 P_R\right) C^{{ \varphi e}}_{f_1 f_2}
\end{dmath*}

\end{minipage}
\end{tabular}

\bigskip

\noindent \begin{tabular}{cp{1mm}l}
\begin{picture}(90,100)(0,50)
\ArrowLine(40,50)(5,50)
\Text(0,50)[r]{$ e^{f_1}$}
\ArrowLine(40,100)(40,50)
\Text(45,95)[l]{$e^{f_2}$}
\DashLine(75,50)(40,50){3}
\Text(80,50)[l]{$h$}
\DashLine(40,0)(40,50){3}
\Text(45,5)[l]{$h$}
\Vertex(40,50){2}
\end{picture}
&&
\begin{minipage}[c]{0.8\linewidth}
\begin{dmath*}
 + \frac{3 i v}{\sqrt2} \left(P_L C^{{e \varphi }*}_{f_2 f_1} + P_R C^{{e \varphi }}_{f_1 f_2}\right)
\end{dmath*}

\end{minipage}
\end{tabular}

\bigskip

\noindent \begin{tabular}{cp{1mm}l}
\begin{picture}(90,100)(0,50)
\ArrowLine(40,50)(5,50)
\Text(0,50)[r]{$\nu^{f_1}$}
\ArrowLine(40,100)(40,50)
\Text(45,95)[l]{$e^{f_2}$}
\DashArrowLine(75,50)(40,50){3}
\Text(80,50)[l]{$G^+$}
\DashLine(40,0)(40,50){3}
\Text(45,5)[l]{$h$}
\Vertex(40,50){2}
\end{picture}
&&
\begin{minipage}[c]{0.8\linewidth}
\begin{dmath*}
+ i {\sqrt2 \over v^2} m_{\nu_{f_1}} U^*_{f_2 f_1}  P_L 
 + i v P_R U_{{g_1} f_1}^* C^{{e \varphi }}_{{g_1} f_2} + i \sqrt2 \left(\slashed{p}_3 P_L - \slashed{p}_4 P_L\right) U_{{g_1} f_1}^* C^{{ \varphi l3}}_{{g_1} f_2}
\end{dmath*}

\end{minipage}
\end{tabular}

\bigskip

\noindent \begin{tabular}{cp{1mm}l}
\begin{picture}(90,100)(0,50)
\ArrowLine(40,50)(5,50)
\Text(0,50)[r]{$\nu^{f_1}$}
\ArrowLine(40,100)(40,50)
\Text(45,95)[l]{$\nu^{f_2}$}
\DashArrowLine(75,50)(40,50){3}
\Text(80,50)[l]{$G^+$}
\DashArrowLine(40,0)(40,50){3}
\Text(45,5)[l]{$G^-$}
\Vertex(40,50){2}
\end{picture}
&&
\begin{minipage}[c]{0.8\linewidth}
\begin{dmath*}
\nolinebreak{+ i \Big[U_{{g_2} f_2} U_{{g_1} f_1}^* C^{{ \varphi l1}}_{{g_1} {g_2}} \left(\slashed{p}_3 P_L - \slashed{p}_4 P_L\right) - U_{{g_1} f_1} U_{{g_2} f_2}^* C^{{ \varphi l1}}_{{g_2} {g_1}} \left(\slashed{p}_3 P_R - \slashed{p}_4 P_R\right)\Big]}
\linebreak
\nolinebreak{+ i \Big[U_{{g_2} f_2} U_{{g_1} f_1}^* C^{{ \varphi l3}}_{{g_1} {g_2}} \left(\slashed{p}_3 P_L - \slashed{p}_4 P_L\right) - U_{{g_1} f_1} U_{{g_2} f_2}^* C^{{ \varphi l3}}_{{g_2} {g_1}} \left(\slashed{p}_3 P_R - \slashed{p}_4 P_R\right)\Big]}
\end{dmath*}

\end{minipage}
\end{tabular}

\bigskip
\noindent \begin{tabular}{cp{1mm}l}
\begin{picture}(90,100)(0,50)
\ArrowLine(40,50)(5,50)
\Text(0,50)[r]{$ \nu^{f_1}$}
\ArrowLine(40,100)(40,50)
\Text(45,95)[l]{$\nu^{f_2}$}
\DashLine(75,50)(40,50){3}
\Text(80,50)[l]{$G^0$}
\DashLine(40,0)(40,50){3}
\Text(45,5)[l]{$G^0$}
\Vertex(40,50){2}
\end{picture}
&&
\begin{minipage}[c]{0.8\linewidth}
\begin{dmath*}
 + \frac{2i}{v^2} \delta_{f_1 f_2} m_{\nu_{f_1}}
\end{dmath*}

\end{minipage}
\end{tabular}

\bigskip
\noindent \begin{tabular}{cp{1mm}l}
\begin{picture}(90,100)(0,50)
\ArrowLine(40,50)(5,50)
\Text(0,50)[r]{$\nu^{f_1}$}
\ArrowLine(40,100)(40,50)
\Text(45,95)[l]{$\nu^{f_2}$}
\DashLine(75,50)(40,50){3}
\Text(80,50)[l]{$G^0$}
\DashLine(40,0)(40,50){3}
\Text(45,5)[l]{$h$}
\Vertex(40,50){2}
\end{picture}
&&
\begin{minipage}[c]{0.8\linewidth}
\begin{dmath*}
-{2\over v^2}\delta_{f_1f_2} m_{\nu_{f_1}}\gamma^5 \linebreak
\nolinebreak{ -  \Big[U_{{g_2} f_2} U_{{g_1} f_1}^* C^{{ \varphi l1}}_{{g_1} {g_2}} \left(\slashed{p}_3 P_L - \slashed{p}_4 P_L\right)-U_{{g_1} f_1} U_{{g_2} f_2}^* C^{{ \varphi l1}}_{{g_2} {g_1}} \left(\slashed{p}_3 P_R - \slashed{p}_4 P_R\right) \Big] }
\linebreak
\nolinebreak{+ \Big[U_{{g_2} f_2}  U_{{g_1} f_1}^* C^{{ \varphi l3}}_{{g_1} {g_2}} \left(\slashed{p}_3 P_L - \slashed{p}_4 P_L\right) - U_{{g_1} f_1}  U_{{g_2} f_2}^* C^{{ \varphi l3}}_{{g_2} {g_1}} \left(\slashed{p}_3 P_R - \slashed{p}_4 P_R\right)\Big]}
\end{dmath*}

\end{minipage}
\end{tabular}

\bigskip

\noindent \begin{tabular}{cp{1mm}l}
\begin{picture}(90,100)(0,50)
\ArrowLine(40,50)(5,50)
\Text(0,50)[r]{$\nu^{f_1}$}
\ArrowLine(40,100)(40,50)
\Text(45,95)[l]{$\nu^{f_2}$}
\DashLine(75,50)(40,50){3}
\Text(80,50)[l]{$h$}
\DashLine(40,0)(40,50){3}
\Text(45,5)[l]{$h$}
\Vertex(40,50){2}
\end{picture}
&&
\begin{minipage}[c]{0.8\linewidth}
\begin{dmath*}
 - \frac{2i}{v^2} \delta_{f_1 f_2} m_{\nu_{f_1}}
\end{dmath*}

\end{minipage}
\end{tabular}

\bigskip

\noindent \begin{tabular}{cp{1mm}l}
\begin{picture}(90,100)(0,50)
\ArrowLine(40,50)(5,50)
\Text(0,50)[r]{$\nu^{f_1}$}
\ArrowLine(40,100)(40,50)
\Text(45,95)[l]{$\nu^{f_2}$}
\DashLine(75,50)(40,50){3}
\Text(80,50)[l]{$h$}
\Photon(40,0)(40,50){3}{4}
\Text(45,5)[l]{$Z^0_{\mu_4}$}
\Vertex(40,50){2}
\end{picture}
&&
\begin{minipage}[c]{0.8\linewidth}
\begin{dmath*}
 \nolinebreak{+ i v \sqrt{{{\bar g}}^2 + {{\bar g'}{}}^2} \Big[U_{{g_2} f_2} U_{{g_1} f_1}^* C^{{ \varphi l1}}_{{g_1} {g_2}} \gamma^{\mu_4} P_L -U_{{g_1} f_1} U_{{g_2} f_2}^* C^{{ \varphi l1}}_{{g_2} {g_1}} \gamma^{\mu_4} P_R\Big]}
 \linebreak
 \nolinebreak{- i v \sqrt{{{\bar g}}^2 + {{\bar g'}{}}^2} \Big[U_{{g_2} f_2} U_{{g_1} f_1}^* C^{{ \varphi l3}}_{{g_1} {g_2}} \gamma^{\mu_4} P_L - U_{{g_1} f_1} U_{{g_2} f_2}^* C^{{ \varphi l3}}_{{g_2} {g_1}} \gamma^{\mu_4} P_R \Big] }
\end{dmath*}

\end{minipage}
\end{tabular}

\bigskip

\noindent \begin{tabular}{cp{1mm}l}
\begin{picture}(90,100)(0,50)
\ArrowLine(40,50)(5,50)
\Text(0,50)[r]{$ e^{f_1}$}
\ArrowLine(40,100)(40,50)
\Text(45,95)[l]{$e^{f_2}$}
\Photon(75,50)(40,50){3}{4}
\Text(80,50)[l]{$A^0_{\mu_3}$}
\DashLine(40,0)(40,50){3}
\Text(45,5)[l]{$h$}
\Vertex(40,50){2}
\end{picture}
&&
\begin{minipage}[c]{0.8\linewidth}
\begin{dmath*}
 + \frac{\sqrt2 {{\bar g'}{}}}{\sqrt{{{\bar g}}^2 + {{\bar g'}{}}^2}} p_3^{\nu } \left(C^{{eW}*}_{f_2 f_1} \sigma^{\mu_3 \nu } P_L  + C^{{eW}}_{f_1 f_2} \sigma^{\mu_3 \nu } P_R \right) - \frac{\sqrt2 {{\bar g}}}{\sqrt{{{\bar g}}^2 + {{\bar g'}{}}^2}} p_3^{\nu } \left(C^{{eB}*}_{f_2 f_1} \sigma^{\mu_3 \nu } P_L  + C^{{eB}}_{f_1 f_2} \sigma^{\mu_3 \nu } P_R \right)
\end{dmath*}

\end{minipage}
\end{tabular}

\bigskip

\noindent \begin{tabular}{cp{1mm}l}
\begin{picture}(90,100)(0,50)
\ArrowLine(40,50)(5,50)
\Text(0,50)[r]{$ e^{f_1}$}
\ArrowLine(40,100)(40,50)
\Text(45,95)[l]{$e^{f_2}$}
\DashLine(75,50)(40,50){3}
\Text(80,50)[l]{$h$}
\Photon(40,0)(40,50){3}{4}
\Text(45,5)[l]{$Z^0_{\mu_4}$}
\Vertex(40,50){2}
\end{picture}
&&
\begin{minipage}[c]{0.8\linewidth}
\begin{dmath*}
 + \frac{\sqrt2 {{\bar g}}}{\sqrt{{{\bar g}}^2 + {{\bar g'}{}}^2}} p_4^{\nu } \left(C^{{eW}*}_{f_2 f_1} \sigma^{\mu_4 \nu } P_L  + C^{{eW}}_{f_1 f_2} \sigma^{\mu_4 \nu } P_R \right) + \frac{\sqrt2 {{\bar g'}{}}}{\sqrt{{{\bar g}}^2 + {{\bar g'}{}}^2}} p_4^{\nu } \left(C^{{eB}*}_{f_2 f_1} \sigma^{\mu_4 \nu } P_L  + C^{{eB}}_{f_1 f_2} \sigma^{\mu_4 \nu } P_R \right) + i v \sqrt{{{\bar g}}^2 + {{\bar g'}{}}^2} C^{{ \varphi l1}}_{f_1 f_2} \gamma^{\mu_4} P_L + i v \sqrt{{{\bar g}}^2 + {{\bar g'}{}}^2} C^{{ \varphi l3}}_{f_1 f_2} \gamma^{\mu_4} P_L + i v \sqrt{{{\bar g}}^2 + {{\bar g'}{}}^2} C^{{ \varphi e}}_{f_1 f_2} \gamma^{\mu_4} P_R
\end{dmath*}

\end{minipage}
\end{tabular}

\bigskip

\noindent \begin{tabular}{cp{1mm}l}
\begin{picture}(90,100)(0,50)
\ArrowLine(40,50)(5,50)
\Text(0,50)[r]{$\nu^{f_1}$}
\ArrowLine(40,100)(40,50)
\Text(45,95)[l]{$\nu^{f_2}$}
\DashArrowLine(75,50)(40,50){3}
\Text(80,50)[l]{$G^-$}
\Photon(40,0)(40,50){3}{4}
\Line(40.,25.)(34.,19.)
\Line(40.,25.)(46.,19.)
\Text(45,5)[l]{$W^+_{\mu_4}$}
\Vertex(40,50){2}
\end{picture}
&&
\begin{minipage}[c]{0.8\linewidth}
\begin{dmath*}
 - i {{\bar g}} v \Big[U_{{g_2} f_2} U_{{g_1} f_1}^* C^{{ \varphi l1}}_{{g_1} {g_2}} \gamma^{\mu_4} P_L - U_{{g_1} f_1} U_{{g_2} f_2}^* C^{{ \varphi l1}}_{{g_2} {g_1}} \gamma^{\mu_4} P_R\Big]
\end{dmath*}

\end{minipage}
\end{tabular}

\bigskip

\noindent \begin{tabular}{cp{1mm}l}
\begin{picture}(90,100)(0,50)
\ArrowLine(40,50)(5,50)
\Text(0,50)[r]{$ e^{f_1}$}
\ArrowLine(40,100)(40,50)
\Text(45,95)[l]{$e^{f_2}$}
\DashArrowLine(75,50)(40,50){3}
\Text(80,50)[l]{$G^-$}
\Photon(40,0)(40,50){3}{4}
\Line(40.,25.)(34.,19.)
\Line(40.,25.)(46.,19.)
\Text(45,5)[l]{$W^+_{\mu_4}$}
\Vertex(40,50){2}
\end{picture}
&&
\begin{minipage}[c]{0.8\linewidth}
\begin{dmath*}
 - 2 \sqrt2 p_4^{\nu } C^{{eW}*}_{f_2 f_1} \sigma^{\mu_4 \nu } P_L  - i {{\bar g}} v C^{{ \varphi l1}}_{f_1 f_2} \gamma^{\mu_4} P_L - i {{\bar g}} v C^{{ \varphi e}}_{f_1 f_2} \gamma^{\mu_4} P_R
\end{dmath*}

\end{minipage}
\end{tabular}

\bigskip

\noindent \begin{tabular}{cp{1mm}l}
\begin{picture}(90,100)(0,50)
\ArrowLine(40,50)(5,50)
\Text(0,50)[r]{$\nu^{f_1}$}
\ArrowLine(40,100)(40,50)
\Text(45,95)[l]{$e^{f_2}$}
\DashLine(75,50)(40,50){3}
\Text(80,50)[l]{$G^0$}
\DashArrowLine(40,0)(40,50){3}
\Text(45,5)[l]{$G^+$}
\Vertex(40,50){2}
\end{picture}
&&
\begin{minipage}[c]{0.8\linewidth}
\begin{dmath*}
- {\sqrt2 \over v^2} m_{\nu_{f_1}} U^*_{f_2 f_1} P_L
 - \sqrt2 \left(\slashed{p}_3 P_L - \slashed{p}_4 P_L\right) U_{{g_1} f_1}^* C^{{ \varphi l3}}_{{g_1} f_2}
\end{dmath*}

\end{minipage}
\end{tabular}

\bigskip

\noindent \begin{tabular}{cp{1mm}l}
\begin{picture}(90,100)(0,50)
\ArrowLine(40,50)(5,50)
\Text(0,50)[r]{$ e^{f_1}$}
\ArrowLine(40,100)(40,50)
\Text(45,95)[l]{$\nu^{f_2}$}
\Photon(75,50)(40,50){3}{4}
\Text(80,50)[l]{$A^0_{\mu_3}$}
\DashArrowLine(40,0)(40,50){3}
\Text(45,5)[l]{$G^-$}
\Vertex(40,50){2}
\end{picture}
&&
\begin{minipage}[c]{0.8\linewidth}
\begin{dmath*}
 - \frac{2 {{\bar g'}{}}}{\sqrt{{{\bar g}}^2 + {{\bar g'}{}}^2}} p_3^{\nu } U_{{g_1} f_2} \sigma^{\mu_3 \nu } P_L  C^{{eW}*}_{{g_1} f_1} - \frac{2 {{\bar g}}}{\sqrt{{{\bar g}}^2 + {{\bar g'}{}}^2}} p_3^{\nu } U_{{g_1} f_2} \sigma^{\mu_3 \nu } P_L  C^{{eB}*}_{{g_1} f_1} - \frac{i \sqrt2 {{\bar g}} {{\bar g'}{}} v}{\sqrt{{{\bar g}}^2 + {{\bar g'}{}}^2}} U_{{g_1} f_2} C^{{ \varphi l3}}_{f_1 {g_1}} \gamma^{\mu_3} P_L
\end{dmath*}

\end{minipage}
\end{tabular}

\bigskip

\noindent \begin{tabular}{cp{1mm}l}
\begin{picture}(90,100)(0,50)
\ArrowLine(40,50)(5,50)
\Text(0,50)[r]{$ e^{f_1}$}
\ArrowLine(40,100)(40,50)
\Text(45,95)[l]{$\nu^{f_2}$}
\DashArrowLine(75,50)(40,50){3}
\Text(80,50)[l]{$G^-$}
\Photon(40,0)(40,50){3}{4}
\Text(45,5)[l]{$Z^0_{\mu_4}$}
\Vertex(40,50){2}
\end{picture}
&&
\begin{minipage}[c]{0.8\linewidth}
\begin{dmath*}
 - \frac{2 {{\bar g}}}{\sqrt{{{\bar g}}^2 + {{\bar g'}{}}^2}} p_4^{\nu } U_{{g_1} f_2} \sigma^{\mu_4 \nu } P_L  C^{{eW}*}_{{g_1} f_1} + \frac{2 {{\bar g'}{}}}{\sqrt{{{\bar g}}^2 + {{\bar g'}{}}^2}} p_4^{\nu } U_{{g_1} f_2} \sigma^{\mu_4 \nu } P_L  C^{{eB}*}_{{g_1} f_1} + \frac{i \sqrt2 {{\bar g'}{}}^2 v}{\sqrt{{{\bar g}}^2 + {{\bar g'}{}}^2}} U_{{g_1} f_2} C^{{ \varphi l3}}_{f_1 {g_1}} \gamma^{\mu_4} P_L
\end{dmath*}

\end{minipage}
\end{tabular}

\bigskip

\noindent \begin{tabular}{cp{1mm}l}
\begin{picture}(90,100)(0,50)
\ArrowLine(40,50)(5,50)
\Text(0,50)[r]{$\nu^{f_1}$}
\ArrowLine(40,100)(40,50)
\Text(45,95)[l]{$e^{f_2}$}
\DashLine(75,50)(40,50){3}
\Text(80,50)[l]{$h$}
\Photon(40,0)(40,50){3}{4}
\Line(40.,25.)(34.,19.)
\Line(40.,25.)(46.,19.)
\Text(45,5)[l]{$W^+_{\mu_4}$}
\Vertex(40,50){2}
\end{picture}
&&
\begin{minipage}[c]{0.8\linewidth}
\begin{dmath*}
 - 2 p_4^{\nu } U_{{g_1} f_1}^* C^{{eW}}_{{g_1} f_2} \sigma^{\mu_4 \nu } P_R  - i \sqrt2 {{\bar g}} v U_{{g_1} f_1}^* C^{{ \varphi l3}}_{{g_1} f_2} \gamma^{\mu_4} P_L
\end{dmath*}

\end{minipage}
\end{tabular}

\bigskip

\noindent \begin{tabular}{cp{1mm}l}
\begin{picture}(90,100)(0,50)
\ArrowLine(40,50)(5,50)
\Text(0,50)[r]{$ e^{f_1}$}
\ArrowLine(40,100)(40,50)
\Text(45,95)[l]{$\nu^{f_2}$}
\DashLine(75,50)(40,50){3}
\Text(80,50)[l]{$G^0$}
\Photon(40,0)(40,50){3}{4}
\Line(40.,25.)(34.,19.)
\Line(40.,25.)(46.,19.)
\Text(45,5)[l]{$W^-_{\mu_4}$}
\Vertex(40,50){2}
\end{picture}
&&
\begin{minipage}[c]{0.8\linewidth}
\begin{dmath*}
 + 2 i p_4^{\nu } U_{{g_1} f_2} \sigma^{\mu_4 \nu } P_L  C^{{eW}*}_{{g_1} f_1}
\end{dmath*}

\end{minipage}
\end{tabular}

\bigskip

\noindent \begin{tabular}{cp{1mm}l}
\begin{picture}(90,100)(0,50)
\ArrowLine(40,50)(5,50)
\Text(0,50)[r]{$ e^{f_1}$}
\ArrowLine(40,100)(40,50)
\Text(45,95)[l]{$e^{f_2}$}
\Photon(75,50)(40,50){3}{4}
\Text(80,50)[l]{$A^0_{\mu_3}$}
\DashLine(40,0)(40,50){3}
\Text(45,5)[l]{$G^0$}
\Vertex(40,50){2}
\end{picture}
&&
\begin{minipage}[c]{0.8\linewidth}
\begin{dmath*}
 - \frac{i \sqrt2 {{\bar g'}{}}}{\sqrt{{{\bar g}}^2 + {{\bar g'}{}}^2}} p_3^{\nu } \left(C^{{eW}*}_{f_2 f_1} \sigma^{\mu_3 \nu } P_L  - C^{{eW}}_{f_1 f_2} \sigma^{\mu_3 \nu } P_R \right) + \frac{i \sqrt2 {{\bar g}}}{\sqrt{{{\bar g}}^2 + {{\bar g'}{}}^2}} p_3^{\nu } \left(C^{{eB}*}_{f_2 f_1} \sigma^{\mu_3 \nu } P_L  - C^{{eB}}_{f_1 f_2} \sigma^{\mu_3 \nu } P_R \right)
\end{dmath*}

\end{minipage}
\end{tabular}

\bigskip

\noindent \begin{tabular}{cp{1mm}l}
\begin{picture}(90,100)(0,50)
\ArrowLine(40,50)(5,50)
\Text(0,50)[r]{$ e^{f_1}$}
\ArrowLine(40,100)(40,50)
\Text(45,95)[l]{$e^{f_2}$}
\DashLine(75,50)(40,50){3}
\Text(80,50)[l]{$G^0$}
\Photon(40,0)(40,50){3}{4}
\Text(45,5)[l]{$Z^0_{\mu_4}$}
\Vertex(40,50){2}
\end{picture}
&&
\begin{minipage}[c]{0.8\linewidth}
\begin{dmath*}
 - \frac{i \sqrt2 {{\bar g}}}{\sqrt{{{\bar g}}^2 + {{\bar g'}{}}^2}} p_4^{\nu } \left(C^{{eW}*}_{f_2 f_1} \sigma^{\mu_4 \nu } P_L  - C^{{eW}}_{f_1 f_2} \sigma^{\mu_4 \nu } P_R \right) - \frac{i \sqrt2 {{\bar g'}{}}}{\sqrt{{{\bar g}}^2 + {{\bar g'}{}}^2}} p_4^{\nu } \left(C^{{eB}*}_{f_2 f_1} \sigma^{\mu_4 \nu } P_L  - C^{{eB}}_{f_1 f_2} \sigma^{\mu_4 \nu } P_R \right)
\end{dmath*}

\end{minipage}
\end{tabular}

\bigskip

\noindent \begin{tabular}{cp{1mm}l}
\begin{picture}(90,100)(0,50)
\ArrowLine(40,50)(5,50)
\Text(0,50)[r]{$ e^{f_1}$}
\ArrowLine(20,95)(40,50)
\Text(15,90)[r]{$e^{f_2}$}
\DashLine(60,95)(40,50){3}
\Text(65,90)[l]{$G^0$}
\DashArrowLine(75,50)(40,50){3}
\Text(80,50)[l]{$G^+$}
\DashArrowLine(40,0)(40,50){3}
\Text(45,5)[l]{$G^-$}
\Vertex(40,50){2}
\end{picture}
&&
\begin{minipage}[c]{0.8\linewidth}
\begin{dmath*}
 + \frac1{\sqrt2} \left(P_L C^{{e \varphi }*}_{f_2 f_1} - P_R C^{{e \varphi }}_{f_1 f_2}\right)
\end{dmath*}

\end{minipage}
\end{tabular}

\bigskip

\noindent \begin{tabular}{cp{1mm}l}
\begin{picture}(90,100)(0,50)
\ArrowLine(40,50)(5,50)
\Text(0,50)[r]{$ e^{f_1}$}
\ArrowLine(20,95)(40,50)
\Text(15,90)[r]{$e^{f_2}$}
\DashArrowLine(60,95)(40,50){3}
\Text(65,90)[l]{$G^+$}
\DashArrowLine(75,50)(40,50){3}
\Text(80,50)[l]{$G^-$}
\DashLine(40,0)(40,50){3}
\Text(45,5)[l]{$h$}
\Vertex(40,50){2}
\end{picture}
&&
\begin{minipage}[c]{0.8\linewidth}
\begin{dmath*}
 + \frac{i}{\sqrt2} \left(P_L C^{{e \varphi }*}_{f_2 f_1} + P_R C^{{e \varphi }}_{f_1 f_2}\right)
\end{dmath*}

\end{minipage}
\end{tabular}

\bigskip

\noindent \begin{tabular}{cp{1mm}l}
\begin{picture}(90,100)(0,50)
\ArrowLine(40,50)(5,50)
\Text(0,50)[r]{$ e^{f_1}$}
\ArrowLine(20,95)(40,50)
\Text(15,90)[r]{$e^{f_2}$}
\DashLine(60,95)(40,50){3}
\Text(65,90)[l]{$G^0$}
\DashLine(75,50)(40,50){3}
\Text(80,50)[l]{$G^0$}
\DashLine(40,0)(40,50){3}
\Text(45,5)[l]{$G^0$}
\Vertex(40,50){2}
\end{picture}
&&
\begin{minipage}[c]{0.8\linewidth}
\begin{dmath*}
 + \frac3{\sqrt2} \left(P_L C^{{e \varphi }*}_{f_2 f_1} - P_R C^{{e \varphi }}_{f_1 f_2}\right)
\end{dmath*}

\end{minipage}
\end{tabular}

\bigskip

\noindent \begin{tabular}{cp{1mm}l}
\begin{picture}(90,100)(0,50)
\ArrowLine(40,50)(5,50)
\Text(0,50)[r]{$ e^{f_1}$}
\ArrowLine(20,95)(40,50)
\Text(15,90)[r]{$e^{f_2}$}
\DashLine(60,95)(40,50){3}
\Text(65,90)[l]{$G^0$}
\DashLine(75,50)(40,50){3}
\Text(80,50)[l]{$G^0$}
\DashLine(40,0)(40,50){3}
\Text(45,5)[l]{$h$}
\Vertex(40,50){2}
\end{picture}
&&
\begin{minipage}[c]{0.8\linewidth}
\begin{dmath*}
 + \frac{i}{\sqrt2} \left(P_L C^{{e \varphi }*}_{f_2 f_1} + P_R C^{{e \varphi }}_{f_1 f_2}\right)
\end{dmath*}

\end{minipage}
\end{tabular}

\bigskip

\noindent \begin{tabular}{cp{1mm}l}
\begin{picture}(90,100)(0,50)
\ArrowLine(40,50)(5,50)
\Text(0,50)[r]{$ e^{f_1}$}
\ArrowLine(20,95)(40,50)
\Text(15,90)[r]{$e^{f_2}$}
\DashLine(60,95)(40,50){3}
\Text(65,90)[l]{$G^0$}
\DashLine(75,50)(40,50){3}
\Text(80,50)[l]{$h$}
\DashLine(40,0)(40,50){3}
\Text(45,5)[l]{$h$}
\Vertex(40,50){2}
\end{picture}
&&
\begin{minipage}[c]{0.8\linewidth}
\begin{dmath*}
 + \frac1{\sqrt2} \left(P_L C^{{e \varphi }*}_{f_2 f_1} - P_R C^{{e \varphi }}_{f_1 f_2}\right)
\end{dmath*}

\end{minipage}
\end{tabular}

\bigskip

\noindent \begin{tabular}{cp{1mm}l}
\begin{picture}(90,100)(0,50)
\ArrowLine(40,50)(5,50)
\Text(0,50)[r]{$ e^{f_1}$}
\ArrowLine(20,95)(40,50)
\Text(15,90)[r]{$e^{f_2}$}
\DashLine(60,95)(40,50){3}
\Text(65,90)[l]{$h$}
\DashLine(75,50)(40,50){3}
\Text(80,50)[l]{$h$}
\DashLine(40,0)(40,50){3}
\Text(45,5)[l]{$h$}
\Vertex(40,50){2}
\end{picture}
&&
\begin{minipage}[c]{0.8\linewidth}
\begin{dmath*}
 + \frac{3 i}{\sqrt2} \left(P_L C^{{e \varphi }*}_{f_2 f_1} + P_R C^{{e \varphi }}_{f_1 f_2}\right)
\end{dmath*}

\end{minipage}
\end{tabular}

\bigskip

\noindent \begin{tabular}{cp{1mm}l}
\begin{picture}(90,100)(0,50)
\ArrowLine(40,50)(5,50)
\Text(0,50)[r]{$\nu^{f_1}$}
\ArrowLine(20,95)(40,50)
\Text(15,90)[r]{$e^{f_2}$}
\DashArrowLine(60,95)(40,50){3}
\Text(65,90)[l]{$G^+$}
\DashArrowLine(75,50)(40,50){3}
\Text(80,50)[l]{$G^+$}
\DashArrowLine(40,0)(40,50){3}
\Text(45,5)[l]{$G^-$}
\Vertex(40,50){2}
\end{picture}
&&
\begin{minipage}[c]{0.8\linewidth}
\begin{dmath*}
 + 2 i P_R U_{{g_1} f_1}^* C^{{e \varphi }}_{{g_1} f_2}
\end{dmath*}

\end{minipage}
\end{tabular}

\bigskip

\noindent \begin{tabular}{cp{1mm}l}
\begin{picture}(90,100)(0,50)
\ArrowLine(40,50)(5,50)
\Text(0,50)[r]{$\nu^{f_1}$}
\ArrowLine(20,95)(40,50)
\Text(15,90)[r]{$e^{f_2}$}
\DashLine(60,95)(40,50){3}
\Text(65,90)[l]{$G^0$}
\DashLine(75,50)(40,50){3}
\Text(80,50)[l]{$G^0$}
\DashArrowLine(40,0)(40,50){3}
\Text(45,5)[l]{$G^+$}
\Vertex(40,50){2}
\end{picture}
&&
\begin{minipage}[c]{0.8\linewidth}
\begin{dmath*}
 + i P_R U_{{g_1} f_1}^* C^{{e \varphi }}_{{g_1} f_2}
\end{dmath*}

\end{minipage}
\end{tabular}

\bigskip

\noindent \begin{tabular}{cp{1mm}l}
\begin{picture}(90,100)(0,50)
\ArrowLine(40,50)(5,50)
\Text(0,50)[r]{$\nu^{f_1}$}
\ArrowLine(20,95)(40,50)
\Text(15,90)[r]{$e^{f_2}$}
\DashArrowLine(60,95)(40,50){3}
\Text(65,90)[l]{$G^+$}
\DashLine(75,50)(40,50){3}
\Text(80,50)[l]{$h$}
\DashLine(40,0)(40,50){3}
\Text(45,5)[l]{$h$}
\Vertex(40,50){2}
\end{picture}
&&
\begin{minipage}[c]{0.8\linewidth}
\begin{dmath*}
 + i P_R U_{{g_1} f_1}^* C^{{e \varphi }}_{{g_1} f_2}
\end{dmath*}

\end{minipage}
\end{tabular}

\bigskip

\noindent \begin{tabular}{cp{1mm}l}
\begin{picture}(90,100)(0,50)
\ArrowLine(40,50)(5,50)
\Text(0,50)[r]{$\nu^{f_1}$}
\ArrowLine(20,95)(40,50)
\Text(15,90)[r]{$\nu^{f_2}$}
\Photon(60,95)(40,50){3}{4}
\Text(65,90)[l]{$A^0_{\mu_3}$}
\DashArrowLine(75,50)(40,50){3}
\Text(80,50)[l]{$G^+$}
\DashArrowLine(40,0)(40,50){3}
\Text(45,5)[l]{$G^-$}
\Vertex(40,50){2}
\end{picture}
&&
\begin{minipage}[c]{0.8\linewidth}
\begin{dmath*}
\nolinebreak{ - \frac{2 i {{\bar g}} {{\bar g'}{}}}{\sqrt{{{\bar g}}^2 + {{\bar g'}{}}^2}} \Big[U_{{g_2} f_2} U_{{g_1} f_1}^* C^{{ \varphi l1}}_{{g_1} {g_2}} \gamma^{\mu_3} P_L -
 U_{{g_1} f_1} U_{{g_2} f_2}^* C^{{ \varphi l1}}_{{g_2} {g_1}} \gamma^{\mu_3} P_R \Big]}
 \linebreak
 \nolinebreak{ - \frac{2 i {{\bar g}} {{\bar g'}{}}}{\sqrt{{{\bar g}}^2 + {{\bar g'}{}}^2}} \Big[U_{{g_2} f_2} U_{{g_1} f_1}^* C^{{ \varphi l3}}_{{g_1} {g_2}} \gamma^{\mu_3} P_L - U_{{g_1} f_1} U_{{g_2} f_2}^* C^{{ \varphi l3}}_{{g_2} {g_1}} \gamma^{\mu_3} P_R\Big]}
\end{dmath*}

\end{minipage}
\end{tabular}

\bigskip

\noindent \begin{tabular}{cp{1mm}l}
\begin{picture}(90,100)(0,50)
\ArrowLine(40,50)(5,50)
\Text(0,50)[r]{$\nu^{f_1}$}
\ArrowLine(20,95)(40,50)
\Text(15,90)[r]{$\nu^{f_2}$}
\DashArrowLine(60,95)(40,50){3}
\Text(65,90)[l]{$G^+$}
\DashArrowLine(75,50)(40,50){3}
\Text(80,50)[l]{$G^-$}
\Photon(40,0)(40,50){3}{4}
\Text(45,5)[l]{$Z^0_{\mu_5}$}
\Vertex(40,50){2}
\end{picture}
&&
\begin{minipage}[c]{0.8\linewidth}
\begin{dmath*}
 \nolinebreak{+ \frac{i \left({{\bar g'}{}}^2 - {{\bar g}}^2\right)}{\sqrt{{{\bar g}}^2 + {{\bar g'}{}}^2}} \Big[U_{{g_2} f_2} U_{{g_1} f_1}^* C^{{ \varphi l1}}_{{g_1} {g_2}} \gamma^{\mu_5} P_L - U_{{g_1} f_1} U_{{g_2} f_2}^* C^{{ \varphi l1}}_{{g_2} {g_1}} \gamma^{\mu_5} P_R \Big]}
 \linebreak
 \nolinebreak{ + \frac{i \left({{\bar g'}{}}^2 - {{\bar g}}^2\right)}{\sqrt{{{\bar g}}^2 + {{\bar g'}{}}^2}} \Big[U_{{g_2} f_2} U_{{g_1} f_1}^* C^{{ \varphi l3}}_{{g_1} {g_2}} \gamma^{\mu_5} P_L-U_{{g_1} f_1} U_{{g_2} f_2}^* C^{{ \varphi l3}}_{{g_2} {g_1}} \gamma^{\mu_5} P_R\Big]}
\end{dmath*}

\end{minipage}
\end{tabular}

\bigskip

\noindent \begin{tabular}{cp{1mm}l}
\begin{picture}(90,100)(0,50)
\ArrowLine(40,50)(5,50)
\Text(0,50)[r]{$ e^{f_1}$}
\ArrowLine(20,95)(40,50)
\Text(15,90)[r]{$e^{f_2}$}
\Photon(60,95)(40,50){3}{4}
\Text(65,90)[l]{$A^0_{\mu_3}$}
\DashArrowLine(75,50)(40,50){3}
\Text(80,50)[l]{$G^+$}
\DashArrowLine(40,0)(40,50){3}
\Text(45,5)[l]{$G^-$}
\Vertex(40,50){2}
\end{picture}
&&
\begin{minipage}[c]{0.8\linewidth}
\begin{dmath*}
 - \frac{2 i {{\bar g}} {{\bar g'}{}}}{\sqrt{{{\bar g}}^2 + {{\bar g'}{}}^2}} C^{{ \varphi l1}}_{f_1 f_2} \gamma^{\mu_3} P_L + \frac{2 i {{\bar g}} {{\bar g'}{}}}{\sqrt{{{\bar g}}^2 + {{\bar g'}{}}^2}} C^{{ \varphi l3}}_{f_1 f_2} \gamma^{\mu_3} P_L - \frac{2 i {{\bar g}} {{\bar g'}{}}}{\sqrt{{{\bar g}}^2 + {{\bar g'}{}}^2}} C^{{ \varphi e}}_{f_1 f_2} \gamma^{\mu_3} P_R
\end{dmath*}

\end{minipage}
\end{tabular}

\bigskip

\noindent \begin{tabular}{cp{1mm}l}
\begin{picture}(90,100)(0,50)
\ArrowLine(40,50)(5,50)
\Text(0,50)[r]{$ e^{f_1}$}
\ArrowLine(20,95)(40,50)
\Text(15,90)[r]{$e^{f_2}$}
\DashArrowLine(60,95)(40,50){3}
\Text(65,90)[l]{$G^+$}
\DashArrowLine(75,50)(40,50){3}
\Text(80,50)[l]{$G^-$}
\Photon(40,0)(40,50){3}{4}
\Text(45,5)[l]{$Z^0_{\mu_5}$}
\Vertex(40,50){2}
\end{picture}
&&
\begin{minipage}[c]{0.8\linewidth}
\begin{dmath*}
 + \frac{i \left({{\bar g'}{}}^2 - {{\bar g}}^2\right)}{\sqrt{{{\bar g}}^2 + {{\bar g'}{}}^2}} C^{{ \varphi l1}}_{f_1 f_2} \gamma^{\mu_5} P_L - \frac{i \left({{\bar g'}{}}^2 - {{\bar g}}^2\right)}{\sqrt{{{\bar g}}^2 + {{\bar g'}{}}^2}} C^{{ \varphi l3}}_{f_1 f_2} \gamma^{\mu_5} P_L + \frac{i \left({{\bar g'}{}}^2 - {{\bar g}}^2\right)}{\sqrt{{{\bar g}}^2 + {{\bar g'}{}}^2}} C^{{ \varphi e}}_{f_1 f_2} \gamma^{\mu_5} P_R
\end{dmath*}

\end{minipage}
\end{tabular}

\bigskip

\noindent \begin{tabular}{cp{1mm}l}
\begin{picture}(90,100)(0,50)
\ArrowLine(40,50)(5,50)
\Text(0,50)[r]{$\nu^{f_1}$}
\ArrowLine(20,95)(40,50)
\Text(15,90)[r]{$\nu^{f_2}$}
\DashLine(60,95)(40,50){3}
\Text(65,90)[l]{$G^0$}
\DashLine(75,50)(40,50){3}
\Text(80,50)[l]{$G^0$}
\Photon(40,0)(40,50){3}{4}
\Text(45,5)[l]{$Z^0_{\mu_5}$}
\Vertex(40,50){2}
\end{picture}
&&
\begin{minipage}[c]{0.8\linewidth}
\begin{dmath*}
\nolinebreak{ + i \sqrt{{{\bar g}}^2 + {{\bar g'}{}}^2} \Big[U_{{g_2} f_2} U_{{g_1} f_1}^* C^{{ \varphi l1}}_{{g_1} {g_2}} \gamma^{\mu_5} P_L-U_{{g_1} f_1} U_{{g_2} f_2}^* C^{{ \varphi l1}}_{{g_2} {g_1}} \gamma^{\mu_5} P_R\Big]}
\linebreak
\nolinebreak{ - i \sqrt{{{\bar g}}^2 + {{\bar g'}{}}^2} \Big[U_{{g_2} f_2} U_{{g_1} f_1}^* C^{{ \varphi l3}}_{{g_1} {g_2}} \gamma^{\mu_5} P_L-U_{{g_1} f_1} U_{{g_2} f_2}^* C^{{ \varphi l3}}_{{g_2} {g_1}} \gamma^{\mu_5} P_R\Big]}
\end{dmath*}

\end{minipage}
\end{tabular}

\bigskip

\noindent \begin{tabular}{cp{1mm}l}
\begin{picture}(90,100)(0,50)
\ArrowLine(40,50)(5,50)
\Text(0,50)[r]{$\nu^{f_1}$}
\ArrowLine(20,95)(40,50)
\Text(15,90)[r]{$\nu^{f_2}$}
\DashLine(60,95)(40,50){3}
\Text(65,90)[l]{$h$}
\DashLine(75,50)(40,50){3}
\Text(80,50)[l]{$h$}
\Photon(40,0)(40,50){3}{4}
\Text(45,5)[l]{$Z^0_{\mu_5}$}
\Vertex(40,50){2}
\end{picture}
&&
\begin{minipage}[c]{0.8\linewidth}
\begin{dmath*}
 \nolinebreak{+ i \sqrt{{{\bar g}}^2 + {{\bar g'}{}}^2} \Big[U_{{g_2} f_2} U_{{g_1} f_1}^* C^{{ \varphi l1}}_{{g_1} {g_2}} \gamma^{\mu_5} P_L - U_{{g_1} f_1} U_{{g_2} f_2}^* C^{{ \varphi l1}}_{{g_2} {g_1}} \gamma^{\mu_5} P_R\Big]}
 \linebreak
 \nolinebreak{- i \sqrt{{{\bar g}}^2 + {{\bar g'}{}}^2} \Big[U_{{g_2} f_2} U_{{g_1} f_1}^* C^{{ \varphi l3}}_{{g_1} {g_2}} \gamma^{\mu_5} P_L-U_{{g_1} f_1} U_{{g_2} f_2}^* C^{{ \varphi l3}}_{{g_2} {g_1}} \gamma^{\mu_5} P_R\Big]}
\end{dmath*}

\end{minipage}
\end{tabular}

\bigskip

\noindent \begin{tabular}{cp{1mm}l}
\begin{picture}(90,100)(0,50)
\ArrowLine(40,50)(5,50)
\Text(0,50)[r]{$ e^{f_1}$}
\ArrowLine(20,95)(40,50)
\Text(15,90)[r]{$e^{f_2}$}
\DashLine(60,95)(40,50){3}
\Text(65,90)[l]{$G^0$}
\DashLine(75,50)(40,50){3}
\Text(80,50)[l]{$G^0$}
\Photon(40,0)(40,50){3}{4}
\Text(45,5)[l]{$Z^0_{\mu_5}$}
\Vertex(40,50){2}
\end{picture}
&&
\begin{minipage}[c]{0.8\linewidth}
\begin{dmath*}
 + i \sqrt{{{\bar g}}^2 + {{\bar g'}{}}^2} C^{{ \varphi l1}}_{f_1 f_2} \gamma^{\mu_5} P_L + i \sqrt{{{\bar g}}^2 + {{\bar g'}{}}^2} C^{{ \varphi l3}}_{f_1 f_2} \gamma^{\mu_5} P_L + i \sqrt{{{\bar g}}^2 + {{\bar g'}{}}^2} C^{{ \varphi e}}_{f_1 f_2} \gamma^{\mu_5} P_R
\end{dmath*}

\end{minipage}
\end{tabular}

\bigskip

\noindent \begin{tabular}{cp{1mm}l}
\begin{picture}(90,100)(0,50)
\ArrowLine(40,50)(5,50)
\Text(0,50)[r]{$ e^{f_1}$}
\ArrowLine(20,95)(40,50)
\Text(15,90)[r]{$e^{f_2}$}
\DashLine(60,95)(40,50){3}
\Text(65,90)[l]{$h$}
\DashLine(75,50)(40,50){3}
\Text(80,50)[l]{$h$}
\Photon(40,0)(40,50){3}{4}
\Text(45,5)[l]{$Z^0_{\mu_5}$}
\Vertex(40,50){2}
\end{picture}
&&
\begin{minipage}[c]{0.8\linewidth}
\begin{dmath*}
 + i \sqrt{{{\bar g}}^2 + {{\bar g'}{}}^2} C^{{ \varphi l1}}_{f_1 f_2} \gamma^{\mu_5} P_L + i \sqrt{{{\bar g}}^2 + {{\bar g'}{}}^2} C^{{ \varphi l3}}_{f_1 f_2} \gamma^{\mu_5} P_L + i \sqrt{{{\bar g}}^2 + {{\bar g'}{}}^2} C^{{ \varphi e}}_{f_1 f_2} \gamma^{\mu_5} P_R
\end{dmath*}

\end{minipage}
\end{tabular}

\bigskip

\noindent \begin{tabular}{cp{1mm}l}
\begin{picture}(90,100)(0,50)
\ArrowLine(40,50)(5,50)
\Text(0,50)[r]{$\nu^{f_1}$}
\ArrowLine(20,95)(40,50)
\Text(15,90)[r]{$\nu^{f_2}$}
\DashLine(60,95)(40,50){3}
\Text(65,90)[l]{$G^0$}
\DashArrowLine(75,50)(40,50){3}
\Text(80,50)[l]{$G^-$}
\Photon(40,0)(40,50){3}{4}
\Line(40.,25.)(34.,19.)
\Line(40.,25.)(46.,19.)
\Text(45,5)[l]{$W^+_{\mu_5}$}
\Vertex(40,50){2}
\end{picture}
&&
\begin{minipage}[c]{0.8\linewidth}
\begin{dmath*}
 + {{\bar g}} \Big[U_{{g_2} f_2} U_{{g_1} f_1}^* C^{{ \varphi l1}}_{{g_1} {g_2}} \gamma^{\mu_5} P_L - U_{{g_1} f_1} U_{{g_2} f_2}^* C^{{ \varphi l1}}_{{g_2} {g_1}} \gamma^{\mu_5} P_R\Big]
\end{dmath*}

\end{minipage}
\end{tabular}

\bigskip

\noindent \begin{tabular}{cp{1mm}l}
\begin{picture}(90,100)(0,50)
\ArrowLine(40,50)(5,50)
\Text(0,50)[r]{$\nu^{f_1}$}
\ArrowLine(20,95)(40,50)
\Text(15,90)[r]{$\nu^{f_2}$}
\DashArrowLine(60,95)(40,50){3}
\Text(65,90)[l]{$G^-$}
\DashLine(75,50)(40,50){3}
\Text(80,50)[l]{$h$}
\Photon(40,0)(40,50){3}{4}
\Line(40.,25.)(34.,19.)
\Line(40.,25.)(46.,19.)
\Text(45,5)[l]{$W^+_{\mu_5}$}
\Vertex(40,50){2}
\end{picture}
&&
\begin{minipage}[c]{0.8\linewidth}
\begin{dmath*}
 - i {{\bar g}}\Big[ U_{{g_2} f_2} U_{{g_1} f_1}^* C^{{ \varphi l1}}_{{g_1} {g_2}} \gamma^{\mu_5} P_L -U_{{g_1} f_1} U_{{g_2} f_2}^* C^{{ \varphi l1}}_{{g_2} {g_1}} \gamma^{\mu_5} P_R \Big]
\end{dmath*}

\end{minipage}
\end{tabular}

\bigskip

\noindent \begin{tabular}{cp{1mm}l}
\begin{picture}(90,100)(0,50)
\ArrowLine(40,50)(5,50)
\Text(0,50)[r]{$ e^{f_1}$}
\ArrowLine(20,95)(40,50)
\Text(15,90)[r]{$e^{f_2}$}
\DashLine(60,95)(40,50){3}
\Text(65,90)[l]{$G^0$}
\DashArrowLine(75,50)(40,50){3}
\Text(80,50)[l]{$G^-$}
\Photon(40,0)(40,50){3}{4}
\Line(40.,25.)(34.,19.)
\Line(40.,25.)(46.,19.)
\Text(45,5)[l]{$W^+_{\mu_5}$}
\Vertex(40,50){2}
\end{picture}
&&
\begin{minipage}[c]{0.8\linewidth}
\begin{dmath*}
 + {{\bar g}} C^{{ \varphi l1}}_{f_1 f_2} \gamma^{\mu_5} P_L + {{\bar g}} C^{{ \varphi e}}_{f_1 f_2} \gamma^{\mu_5} P_R
\end{dmath*}

\end{minipage}
\end{tabular}

\bigskip

\noindent \begin{tabular}{cp{1mm}l}
\begin{picture}(90,100)(0,50)
\ArrowLine(40,50)(5,50)
\Text(0,50)[r]{$ e^{f_1}$}
\ArrowLine(20,95)(40,50)
\Text(15,90)[r]{$e^{f_2}$}
\DashArrowLine(60,95)(40,50){3}
\Text(65,90)[l]{$G^-$}
\DashLine(75,50)(40,50){3}
\Text(80,50)[l]{$h$}
\Photon(40,0)(40,50){3}{4}
\Line(40.,25.)(34.,19.)
\Line(40.,25.)(46.,19.)
\Text(45,5)[l]{$W^+_{\mu_5}$}
\Vertex(40,50){2}
\end{picture}
&&
\begin{minipage}[c]{0.8\linewidth}
\begin{dmath*}
 - i {{\bar g}} C^{{ \varphi l1}}_{f_1 f_2} \gamma^{\mu_5} P_L - i {{\bar g}} C^{{ \varphi e}}_{f_1 f_2} \gamma^{\mu_5} P_R
\end{dmath*}

\end{minipage}
\end{tabular}

\bigskip

\noindent \begin{tabular}{cp{1mm}l}
\begin{picture}(90,100)(0,50)
\ArrowLine(40,50)(5,50)
\Text(0,50)[r]{$ e^{f_1}$}
\ArrowLine(20,95)(40,50)
\Text(15,90)[r]{$\nu^{f_2}$}
\Photon(60,95)(40,50){3}{4}
\Text(65,90)[l]{$A^0_{\mu_3}$}
\DashLine(75,50)(40,50){3}
\Text(80,50)[l]{$G^0$}
\DashArrowLine(40,0)(40,50){3}
\Text(45,5)[l]{$G^-$}
\Vertex(40,50){2}
\end{picture}
&&
\begin{minipage}[c]{0.8\linewidth}
\begin{dmath*}
 + \frac{\sqrt2 {{\bar g}} {{\bar g'}{}}}{\sqrt{{{\bar g}}^2 + {{\bar g'}{}}^2}} U_{{g_1} f_2} C^{{ \varphi l3}}_{f_1 {g_1}} \gamma^{\mu_3} P_L
\end{dmath*}

\end{minipage}
\end{tabular}

\bigskip

\noindent \begin{tabular}{cp{1mm}l}
\begin{picture}(90,100)(0,50)
\ArrowLine(40,50)(5,50)
\Text(0,50)[r]{$ e^{f_1}$}
\ArrowLine(20,95)(40,50)
\Text(15,90)[r]{$\nu^{f_2}$}
\Photon(60,95)(40,50){3}{4}
\Text(65,90)[l]{$A^0_{\mu_3}$}
\DashArrowLine(75,50)(40,50){3}
\Text(80,50)[l]{$G^-$}
\DashLine(40,0)(40,50){3}
\Text(45,5)[l]{$h$}
\Vertex(40,50){2}
\end{picture}
&&
\begin{minipage}[c]{0.8\linewidth}
\begin{dmath*}
 - \frac{i \sqrt2 {{\bar g}} {{\bar g'}{}}}{\sqrt{{{\bar g}}^2 + {{\bar g'}{}}^2}} U_{{g_1} f_2} C^{{ \varphi l3}}_{f_1 {g_1}} \gamma^{\mu_3} P_L
\end{dmath*}

\end{minipage}
\end{tabular}

\bigskip

\noindent \begin{tabular}{cp{1mm}l}
\begin{picture}(90,100)(0,50)
\ArrowLine(40,50)(5,50)
\Text(0,50)[r]{$ e^{f_1}$}
\ArrowLine(20,95)(40,50)
\Text(15,90)[r]{$\nu^{f_2}$}
\DashLine(60,95)(40,50){3}
\Text(65,90)[l]{$G^0$}
\DashArrowLine(75,50)(40,50){3}
\Text(80,50)[l]{$G^-$}
\Photon(40,0)(40,50){3}{4}
\Text(45,5)[l]{$Z^0_{\mu_5}$}
\Vertex(40,50){2}
\end{picture}
&&
\begin{minipage}[c]{0.8\linewidth}
\begin{dmath*}
 - \frac{\sqrt2 {{\bar g'}{}}^2}{\sqrt{{{\bar g}}^2 + {{\bar g'}{}}^2}} U_{{g_1} f_2} C^{{ \varphi l3}}_{f_1 {g_1}} \gamma^{\mu_5} P_L
\end{dmath*}

\end{minipage}
\end{tabular}

\bigskip

\noindent \begin{tabular}{cp{1mm}l}
\begin{picture}(90,100)(0,50)
\ArrowLine(40,50)(5,50)
\Text(0,50)[r]{$ e^{f_1}$}
\ArrowLine(20,95)(40,50)
\Text(15,90)[r]{$\nu^{f_2}$}
\DashArrowLine(60,95)(40,50){3}
\Text(65,90)[l]{$G^-$}
\DashLine(75,50)(40,50){3}
\Text(80,50)[l]{$h$}
\Photon(40,0)(40,50){3}{4}
\Text(45,5)[l]{$Z^0_{\mu_5}$}
\Vertex(40,50){2}
\end{picture}
&&
\begin{minipage}[c]{0.8\linewidth}
\begin{dmath*}
 + \frac{i \sqrt2 {{\bar g'}{}}^2}{\sqrt{{{\bar g}}^2 + {{\bar g'}{}}^2}} U_{{g_1} f_2} C^{{ \varphi l3}}_{f_1 {g_1}} \gamma^{\mu_5} P_L
\end{dmath*}

\end{minipage}
\end{tabular}

\bigskip

\noindent \begin{tabular}{cp{1mm}l}
\begin{picture}(90,100)(0,50)
\ArrowLine(40,50)(5,50)
\Text(0,50)[r]{$\nu^{f_1}$}
\ArrowLine(20,95)(40,50)
\Text(15,90)[r]{$e^{f_2}$}
\DashArrowLine(60,95)(40,50){3}
\Text(65,90)[l]{$G^+$}
\DashArrowLine(75,50)(40,50){3}
\Text(80,50)[l]{$G^-$}
\Photon(40,0)(40,50){3}{4}
\Line(40.,25.)(34.,19.)
\Line(40.,25.)(46.,19.)
\Text(45,5)[l]{$W^+_{\mu_5}$}
\Vertex(40,50){2}
\end{picture}
&&
\begin{minipage}[c]{0.8\linewidth}
\begin{dmath*}
 - i \sqrt2 {{\bar g}} U_{{g_1} f_1}^* C^{{ \varphi l3}}_{{g_1} f_2} \gamma^{\mu_5} P_L
\end{dmath*}

\end{minipage}
\end{tabular}

\bigskip

\noindent \begin{tabular}{cp{1mm}l}
\begin{picture}(90,100)(0,50)
\ArrowLine(40,50)(5,50)
\Text(0,50)[r]{$\nu^{f_1}$}
\ArrowLine(20,95)(40,50)
\Text(15,90)[r]{$e^{f_2}$}
\DashLine(60,95)(40,50){3}
\Text(65,90)[l]{$G^0$}
\DashLine(75,50)(40,50){3}
\Text(80,50)[l]{$G^0$}
\Photon(40,0)(40,50){3}{4}
\Line(40.,25.)(34.,19.)
\Line(40.,25.)(46.,19.)
\Text(45,5)[l]{$W^+_{\mu_5}$}
\Vertex(40,50){2}
\end{picture}
&&
\begin{minipage}[c]{0.8\linewidth}
\begin{dmath*}
 - i \sqrt2 {{\bar g}} U_{{g_1} f_1}^* C^{{ \varphi l3}}_{{g_1} f_2} \gamma^{\mu_5} P_L
\end{dmath*}

\end{minipage}
\end{tabular}

\bigskip

\noindent \begin{tabular}{cp{1mm}l}
\begin{picture}(90,100)(0,50)
\ArrowLine(40,50)(5,50)
\Text(0,50)[r]{$\nu^{f_1}$}
\ArrowLine(20,95)(40,50)
\Text(15,90)[r]{$e^{f_2}$}
\DashLine(60,95)(40,50){3}
\Text(65,90)[l]{$h$}
\DashLine(75,50)(40,50){3}
\Text(80,50)[l]{$h$}
\Photon(40,0)(40,50){3}{4}
\Line(40.,25.)(34.,19.)
\Line(40.,25.)(46.,19.)
\Text(45,5)[l]{$W^+_{\mu_5}$}
\Vertex(40,50){2}
\end{picture}
&&
\begin{minipage}[c]{0.8\linewidth}
\begin{dmath*}
 - i \sqrt2 {{\bar g}} U_{{g_1} f_1}^* C^{{ \varphi l3}}_{{g_1} f_2} \gamma^{\mu_5} P_L
\end{dmath*}

\end{minipage}
\end{tabular}

\bigskip

\noindent \begin{tabular}{cp{1mm}l}
\begin{picture}(90,100)(0,50)
\ArrowLine(40,50)(5,50)
\Text(0,50)[r]{$ e^{f_1}$}
\ArrowLine(20,95)(40,50)
\Text(15,90)[r]{$\nu^{f_2}$}
\DashArrowLine(60,95)(40,50){3}
\Text(65,90)[l]{$G^-$}
\Photon(75,50)(40,50){3}{4}
\Line(57.,50.)(63.,44.)
\Line(57.,50.)(63.,56.)
\Text(80,50)[l]{$W^+_{\mu_4}$}
\Photon(40,0)(40,50){3}{4}
\Line(40.,25.)(34.,19.)
\Line(40.,25.)(46.,19.)
\Text(45,5)[l]{$W^-_{\mu_5}$}
\Vertex(40,50){2}
\end{picture}
&&
\begin{minipage}[c]{0.8\linewidth}
\begin{dmath*}
 - 2 {{\bar g}} U_{{g_1} f_2} \sigma^{\mu_4 \mu_5} P_L  C^{{eW}*}_{{g_1} f_1}
\end{dmath*}

\end{minipage}
\end{tabular}

\bigskip

\noindent \begin{tabular}{cp{1mm}l}
\begin{picture}(90,100)(0,50)
\ArrowLine(40,50)(5,50)
\Text(0,50)[r]{$ e^{f_1}$}
\ArrowLine(20,95)(40,50)
\Text(15,90)[r]{$e^{f_2}$}
\DashLine(60,95)(40,50){3}
\Text(65,90)[l]{$G^0$}
\Photon(75,50)(40,50){3}{4}
\Line(57.,50.)(63.,44.)
\Line(57.,50.)(63.,56.)
\Text(80,50)[l]{$W^+_{\mu_4}$}
\Photon(40,0)(40,50){3}{4}
\Line(40.,25.)(34.,19.)
\Line(40.,25.)(46.,19.)
\Text(45,5)[l]{$W^-_{\mu_5}$}
\Vertex(40,50){2}
\end{picture}
&&
\begin{minipage}[c]{0.8\linewidth}
\begin{dmath*}
 - i \sqrt2 {{\bar g}} \left(\sigma^{\mu_4 \mu_5} P_L  C^{{eW}*}_{f_2 f_1} - C^{{eW}}_{f_1 f_2} \sigma^{\mu_4 \mu_5} P_R \right)
\end{dmath*}

\end{minipage}
\end{tabular}

\bigskip

\noindent \begin{tabular}{cp{1mm}l}
\begin{picture}(90,100)(0,50)
\ArrowLine(40,50)(5,50)
\Text(0,50)[r]{$ e^{f_1}$}
\ArrowLine(20,95)(40,50)
\Text(15,90)[r]{$e^{f_2}$}
\DashLine(60,95)(40,50){3}
\Text(65,90)[l]{$h$}
\Photon(75,50)(40,50){3}{4}
\Line(57.,50.)(63.,44.)
\Line(57.,50.)(63.,56.)
\Text(80,50)[l]{$W^+_{\mu_4}$}
\Photon(40,0)(40,50){3}{4}
\Line(40.,25.)(34.,19.)
\Line(40.,25.)(46.,19.)
\Text(45,5)[l]{$W^-_{\mu_5}$}
\Vertex(40,50){2}
\end{picture}
&&
\begin{minipage}[c]{0.8\linewidth}
\begin{dmath*}
 + \sqrt2 {{\bar g}} \left(\sigma^{\mu_4 \mu_5} P_L  C^{{eW}*}_{f_2 f_1} + C^{{eW}}_{f_1 f_2} \sigma^{\mu_4 \mu_5} P_R \right)
\end{dmath*}

\end{minipage}
\end{tabular}

\bigskip

\noindent \begin{tabular}{cp{1mm}l}
\begin{picture}(90,100)(0,50)
\ArrowLine(40,50)(5,50)
\Text(0,50)[r]{$ e^{f_1}$}
\ArrowLine(20,95)(40,50)
\Text(15,90)[r]{$\nu^{f_2}$}
\Photon(60,95)(40,50){3}{4}
\Text(65,90)[l]{$A^0_{\mu_3}$}
\DashLine(75,50)(40,50){3}
\Text(80,50)[l]{$G^0$}
\Photon(40,0)(40,50){3}{4}
\Line(40.,25.)(34.,19.)
\Line(40.,25.)(46.,19.)
\Text(45,5)[l]{$W^-_{\mu_5}$}
\Vertex(40,50){2}
\end{picture}
&&
\begin{minipage}[c]{0.8\linewidth}
\begin{dmath*}
 - \frac{2 i {{\bar g}} {{\bar g'}{}}}{\sqrt{{{\bar g}}^2 + {{\bar g'}{}}^2}} U_{{g_1} f_2} \sigma^{\mu_3 \mu_5} P_L  C^{{eW}*}_{{g_1} f_1}
\end{dmath*}

\end{minipage}
\end{tabular}

\bigskip

\noindent \begin{tabular}{cp{1mm}l}
\begin{picture}(90,100)(0,50)
\ArrowLine(40,50)(5,50)
\Text(0,50)[r]{$ e^{f_1}$}
\ArrowLine(20,95)(40,50)
\Text(15,90)[r]{$\nu^{f_2}$}
\DashLine(60,95)(40,50){3}
\Text(65,90)[l]{$G^0$}
\Photon(75,50)(40,50){3}{4}
\Line(57.,50.)(63.,44.)
\Line(57.,50.)(63.,56.)
\Text(80,50)[l]{$W^-_{\mu_4}$}
\Photon(40,0)(40,50){3}{4}
\Text(45,5)[l]{$Z^0_{\mu_5}$}
\Vertex(40,50){2}
\end{picture}
&&
\begin{minipage}[c]{0.8\linewidth}
\begin{dmath*}
 + \frac{2 i {{\bar g}}^2}{\sqrt{{{\bar g}}^2 + {{\bar g'}{}}^2}} U_{{g_1} f_2} \sigma^{\mu_4 \mu_5} P_L  C^{{eW}*}_{{g_1} f_1}
\end{dmath*}

\end{minipage}
\end{tabular}

\bigskip

\noindent \begin{tabular}{cp{1mm}l}
\begin{picture}(90,100)(0,50)
\ArrowLine(40,50)(5,50)
\Text(0,50)[r]{$ e^{f_1}$}
\ArrowLine(20,95)(40,50)
\Text(15,90)[r]{$\nu^{f_2}$}
\Photon(60,95)(40,50){3}{4}
\Text(65,90)[l]{$A^0_{\mu_3}$}
\DashLine(75,50)(40,50){3}
\Text(80,50)[l]{$h$}
\Photon(40,0)(40,50){3}{4}
\Line(40.,25.)(34.,19.)
\Line(40.,25.)(46.,19.)
\Text(45,5)[l]{$W^-_{\mu_5}$}
\Vertex(40,50){2}
\end{picture}
&&
\begin{minipage}[c]{0.8\linewidth}
\begin{dmath*}
 + \frac{2 {{\bar g}} {{\bar g'}{}}}{\sqrt{{{\bar g}}^2 + {{\bar g'}{}}^2}} U_{{g_1} f_2} \sigma^{\mu_3 \mu_5} P_L  C^{{eW}*}_{{g_1} f_1}
\end{dmath*}

\end{minipage}
\end{tabular}%

\bigskip

\bigskip
\bigskip
\bigskip

\subsection{Quark--gauge vertices}

\noindent %

\bigskip

\bigskip
\bigskip
\bigskip
\subsection{Quark--Higgs--gauge vertices}

\noindent %

\bigskip

\bigskip
\bigskip
\bigskip
\subsection{Quark-gluon vertices}

\noindent %

\bigskip

\bigskip
\bigskip
\bigskip
\subsection{Higgs-gauge vertices}

\noindent %

\bigskip

\bigskip
\bigskip
\bigskip
\subsection{Gauge-gauge vertices}

\noindent %

\bigskip

\bigskip
\bigskip
\bigskip
\subsection{Higgs-gluon vertices}

\noindent %

\bigskip

\bigskip
\bigskip
\bigskip
\subsection{Gluon-gluon vertices}

\noindent %

\bigskip

\bigskip
\bigskip
\bigskip
\subsection{Four-fermion vertices}

\noindent %

\bigskip

\bigskip
\bigskip
\bigskip

\subsection{Lepton and baryon number violating vertices}

\noindent %

\bigskip

\bigskip
\bigskip
\bigskip

\subsection{Ghost vertices}

\noindent %

\bigskip


\bibliography{EFT}{}
\bibliographystyle{JHEP}                        

\end{document}